\documentclass[useAMS,usenatbib]{mnras}

\usepackage{amssymb}
\usepackage{amsmath}
\usepackage{times}
\usepackage{graphicx}
\usepackage{pgf}

\def\sun {\hbox{M$_{\sun}$}} 
 
\def\psr {PSR~B0823$+$26}
\def\psrb {PSR~B0943$+$10}
\def\psrc {PSR~B1822$-$09}
\def\xmm {{\it XMM-Newton}}

\def\eq{\begin{equation}}
\def\en{\end{equation}}

\def\ie{{\it i.e.}\thinspace}

\title[X-ray/radio mode switching of PSR B0823$+$26]{Discovery of synchronous X-ray \& radio moding of PSR~B0823$+$26}
\author[W. Hermsen et al.]{W. Hermsen$^{1,2}$, L. Kuiper$^{1}$, R. Basu$^{4,7}$, J.W.T. Hessels$^{2,3}$, D. Mitra$^{4,5,7}$, J.M. Rankin$^{2,5}$,
\newauthor   B.W. Stappers$^{6}$, G.A.E. Wright$^{6}$, J.-M. Grie{\ss}meier$^{9,8}$, M. Serylak$^{10,11,8}$, A. Horneffer$^{12}$,  
\newauthor C. Tiburzi$^{12,13}$, W.C.G. Ho$^{14,15}$ \\
$^{1}$SRON Netherlands Institute for Space Research, Sorbonnelaan 2, 3584 CA  Utrecht, The Netherlands.\\
$^{2}$Anton $P$annekoek Institute for Astronomy, University of Amsterdam, Science Park 904, 1098 XH, Amsterdam, The Netherlands.\\
$^{3}$ASTRON, the Netherlands Institute for Radio Astronomy, Postbus 2, 7990 AA, Dwingeloo, The Netherlands.\\
$^{4}$National Centre for Radio Astrophysics, (NCFRA-TIFR), Post Bag 3, Ganeshkhind, Pune University Campus, Pune 411007, India.\\
$^{5}$Physics Department, University of Vermont, Burlington, VT 05405, USA.\\
$^{6}$Jodrell Bank Centre for Astrophysics, School of Physics and Astronomy, University of Manchester, Manchester M13 9PL, UK.\\
$^{7}$Janusz Gil Institute of Astronomy, University of Zielona G\'ora, Lubuska 2, PL-65-265 Zielona G\'ora, Poland.\\
$^{8}$Station de Radioastronomie de Nan\c{c}ay, Observatoire de Paris, CNRS, PSL Research University, Univ. d'Orl\'{e}ans, F-18330 Nan\c{c}ay, France.\\
$^{9}$LPC2E - Universit\'{e} d'Orl\'{e}ans, CNRS, 45071 Orl\'{e}ans, France\\
$^{10}$SKA South Africa, The Park, Park Road, Pinelands 7405, South Africa\\
$^{11}$Department of Physics \& Astronomy, University of the Western Cape, Private Bag X17, Bellville 7535, South Africa.\\
$^{12}$Max-Planck-Institut f\"{u}r Radioastronomie, Auf dem H\"{u}gel 69, 53121 Bonn, Germany.\\
$^{13}$Fakult\"{a}t f\"{u}r Physik, Universit\"{a}t Bielefeld, Postfach 100131, 33501 Bielefeld, Germany.\\
$^{14}$Mathematical Sciences, Physics \& Astronomy and STAG Research Centre, University of Southampton, Southampton, SO17 1BJ, UK\\
$^{15}$Department of Physics and Astronomy, Haverford College, 370 Lancaster Avenue, Haverford, PA 19041, USA
}
\begin{document}

\date{Accepted 2018 July 30. Received 2018 July 09; in original form 2018 May 28.}

\pagerange{\pageref{firstpage}--\pageref{lastpage}} \pubyear{2016}

\maketitle

\label{firstpage}

\begin{abstract}
{Simultaneous observations of \psr\ with ESA's \xmm,  the Giant Metrewave Radio Telescope and international stations of the Low Frequency Array revealed 
synchronous X-ray/radio switching between a radio-bright (B) mode and a radio-quiet (Q) mode. 
During the B mode we detected \psr\  in 0.2$-$2 keV X-rays and discovered pulsed emission with a broad
sinusoidal pulse, lagging the radio main pulse by 0.208$\pm$0.012 in phase, with high pulsed fraction of 70$-$80\%. 
During the Q mode \psr\ was not detected in X-rays (2$\sigma$ upper limit a factor $\sim$9 below the B-mode flux).
The total X-ray spectrum, pulse profile and pulsed fraction can globally be reproduced with 
a magnetized partially ionized hydrogen atmosphere model
with three emission components: a primary small hot spot ($T$$\sim$3.6$\times10^6$ K, $R$$\sim$17 m),
a larger cooler concentric ring ($T$$\sim$1.1$\times10^6$ K, $R$$\sim$280 m)
and an antipodal hot spot ($T$$\sim$1.1$\times10^6 $ K, $R$$\sim$100 m), for the angle between the rotation axis and line of sight direction 
$\sim66^\circ$. The latter is in conflict with the radio derived value of $(84\pm0.7)^\circ$. 
The average X-ray flux within hours-long B-mode intervals
varied by a factor $\pm$20\%, possibly correlated with variations in the frequency and lengths of short radio nulls or short durations of weak emission.
The correlated X-ray/radio moding of \psr\ is 
compared with the anti-correlated moding of \psrb, and the lack of X-ray moding of \psrc. 
We speculate that the X-ray/radio switches of \psr\ are due to variations in the rate of accretion of material from the interstellar medium
through which it is passing.

}
\end{abstract}

\begin{keywords}
stars: neutron --- pulsars: general --- Radio continuum: individual: \psr\ --- X-rays: individual: \psr, \psrb, \psrc\
\end{keywords}

\section{Introduction}

So far, synchronous X-ray and radio mode switching has only been observed for the old and nearly aligned \psrb\ \citep{hermsen2013}.
When this paragon of pulsar-mode switching \citep{suleymanova1984, rankin2006} switches from a radio ``bright'' (B) mode 
to a radio ``quiet'' (Q) mode, the X-ray flux increases by a factor of $\sim$2.4 
and the X-ray pulsed fraction changes in magnitude and as a function of energy \citep{mereghetti2016}. The results are broadly consistent
with the predictions of the partially screened gap model for
rotation-powered pulsar emission (e.g. \citet{szary2015}), but they might also imply global
magnetospheric rearrangements to explain the mode switching, as proposed in studies of intermittent pulsars
\citep{kramer2006, lyne2010, camilo2012, lorimer2012}. It is imperative to find other moding pulsars in which this phenomenon
can be precisely characterised and the results compared with \psrb.

In a first attempt to find another example, we have conducted a similar campaign on the radio mode-switching pulsar \psrc\ \citep{fowler1981, fowler1982}.
This pulsar exhibits mode switching on timescales of minutes, much shorter than the characteristic timescale of hours seen for \psrb.
We found no evidence for simultaneous radio/X-ray mode switching of \psrc\ \citep{hermsen2017}. Though,
we did, surprisingly, find a correlation between the mode durations
and regular modulations in the pulse intensity.  For this pulsar, it
is possible that the very different  geometry between the magnetic and
spin axes (nearly orthogonal) may be hiding any visible X-ray mode changes, or that the physics of the short moding is quite different. 

In this work we report on a new observational campaign targeting the moding pulsar \psr. This pulsar
is one of the brightest radio pulsars in the Northern sky, located at a distance of only $\sim$320 pc. 
It has a period $P$ of $\sim530$ ms, a spin-down age of $\sim 4.9 \times 10^6$ yr and an inferred 
magnetic field of $\sim 9.8 \times 10^{11}$ G. It is by no means an ordinary pulsar, showing main-pulse, inter-pulse and 
post-cursor emission components \citep{backer1973}. The pulsar was found to exhibit nulling (i.e. abrupt cessation and 
re-activation of its radio emission; see e.g. \citet{rathnasree1995}. More recently \citet{young2012}) 
found the nulls to be both short-term ($\sim$ min) as well as unusually long-term ($\sim$ hours or more). 
During bursts in its B mode \psr\ was found to exhibit single-pulse modulations with a repetition period $P_3$$\sim5P$ \citep{backer1973a, weltevrede2006a, weltevrede2007}.
\citet{sobey2015} reported the discovery with LOFAR of a very weak quiet (Q) emission mode during 
these apparent nulling intervals. The transition between the Q mode and B mode 
occurs simultaneously for the main pulse, inter-pulse and post-cursor within about a single rotation of the neutron star. 
From the latter work we estimate the pulsar to be in the Q/null mode for $\sim$30\% of the time and the remaining $\sim$70\% in the B mode.

\psr\ was also detected with {\it XMM-Newton} in X-rays in April 2002 \citep{becker2004}, with an effective EPIC Pn exposure of 33.7 ks. 
The detected source count rate in the Pn CCD was $3.5 \times 10^{-3}$ counts s$^{-1}$ for energies 0.3$-$10 keV, or an X-ray luminosity (0.5$-$10 keV) of $\sim 1.0 \times 10^{29}$ erg s$^{-1}$.
Furthermore, there was a 2.2$\sigma$ indication for the detection 
of a pulsed signal. Considerably more exposure is apparently required to confirm the indication for pulsed emission, as well as for significantly 
constraining the spectral parameters of  the X-ray emission. 

We carried out an X-ray/radio campaign on \psr\ with 6 $\times$ 25 ks of {\it XMM-Newton} observations, and simultaneous radio observations primarily with the 
GMRT at 325 MHz, supported by LOFAR
UK, FR, SE and DE International stations in standalone mode at 150 MHz. From Lovell Telescope radio monitoring of
\psr\ we derived an ephemeris to facilitate phase folding of the arrival times of the X-ray events.

In Section 2, we present the radio observations and how we defined the radio modes. In the subsequent sections, we present the {\it XMM-Newton} X-ray observations
(Section 3), the spatial analysis of the sky maps for the six X-ray observations, leading to the discovery of synchronous 
X-ray/radio mode switching (Section 4), the first detection of the X-ray pulsed signal from \psr\ (Section 5), followed by the spectral characterisation of the total and
pulsed emissions (Section 6). In Section 7, we summarise our findings, followed in Section 8 by a discussion of them in comparison with the results from the previous 
campaigns on \psrb\ and \psrc. Finally, our overall conclusions are presented in Section 10.

\section{Radio observations and mode determinations}
\label{modes}
Due to orbital viewing constraints of \xmm, simultaneous radio coverage of our X-ray observations of \psr\ in 2017 could mainly be realised with the GMRT,
with support by LOFAR international stations for the last part of each assigned orbit of the space craft. In those cases when there was a mismatch in the scheduled
\xmm\ / GMRT observations, or when the GMRT encountered operational problems, the LOFAR observations could provide the required information for our 
analysis.

\begin{table}
\caption{Radio observations of \psr\ in April 2017 with the GMRT and the LOFAR DE601 International Telescope (Effelsberg),
together covering most of the simultaneous observations with \xmm\ (Table~\ref{table_xmm_obs}). For definitions of the radio
Bright (B), Null, Quiet (Q) and Q-bright mode, see text.}
\label{table_radio_obs}
\begin{center}
\begin{tabular}{c c c c c c}
\hline
Telesc.    & Date &  Freq. & Start (UT) & End (UT) & Radio \\
            &    (yyyy-mm-dd)  & (MHz)&  (hh:mm)        &    (hh:mm)  & mode   \\
\hline
GMRT &   &   &   &   &   \\   
 & 2017-04-20 & 325  &  08:23  &   16:30 & B    \\
 & 2017-04-22 & 325 & 08:14   &  13:20 & B   \\
  & 2017-04-24 & 325 &08:06  &  09:03 & Null \\
 & 2017-04-24 & 325 &09:03  &  09:10 & Q-bright \\
 & 2017-04-24 & 325 & 09:10  &  16:24 &  Q \\
    & 2017-04-26 & 325 &12:30  &  16:30 & B   \\
  & 2017-04-28 & 325 & 09:26  &  16:05 & B  \\
  & 2017-04-30 & 325 & 09:03  & 12:59 & B   \\
  LOFAR & & & & & \\
  & 2017-04-20 & 150  &  12:46  &   22:49 & B    \\
 & 2017-04-22 & 150 & 12:38   &  22:44 & B   \\
  & 2017-04-24 & 150 & 12:45  &  17:00 & Q\\
 & 2017-04-24 & 150 &17:00  &  22:39 & B \\
 & 2017-04-26 & 150 & 12:22  &  22:26 &  B \\
    & 2017-04-28 & 150 &12:14  &  21:48 & B   \\
  & 2017-04-28 & 150 & 21:48  &  22:18 & Q  \\
  & 2017-04-30 & 150 & 15:27  & 22:07 & B   \\
\hline \\
\end{tabular}
\end{center}
\end{table}

\subsection{GMRT} 
Simultaneous radio/X-ray observations of \psr\ were carried out at the GMRT for six observing
sessions of $\sim$7.5 h each in 2017 April at 325 MHz (see Table~\ref{table_radio_obs}). 
The GMRT is a multi-element
aperture synthesis telescope \citep{swarup1991}.  The observations discussed here were done in a similar
manner as discussed in \citep{hermsen2017}. To maximize the sensitivity for single pulse detection in our
study, we used approximately twenty antennas in the phased array, including all the available central
square antennas and the two nearest arm antennas.
The observing strategy was to do ``phasing'' of the array using a strong
nearby phase calibrator and then observe the pulsar
in an interlaced manner. To avoid dephasing due for long
observations the phasing of the array was done every 1.5 hours,
which however caused missing data in the pulsar time sequence for a few minutes.
We used a total bandwidth
of about 33 MHz in the frequency range 306$-$339 MHz,
spread over 256 channels. The data were recorded in
the filter-bank format with time resolution of 0.122 ms, which was subsequently
averaged to 0.488 ms.

The observing sessions were scheduled such that there was substantial overlap with the \xmm\ observations. Generally,
GMRT commenced the observations later than \xmm\ by 18 minutes up to $\sim$1 hour (compare Table~\ref{table_radio_obs} with Table~\ref{table_xmm_obs}). On April 26
the GMRT encountered problems and started 6 hours 46 minutes later than \xmm, unfortunately. On two occasion, April 22 and 30, 
GMRT stopped observing earlier than \xmm, by 2 hours 24 minutes and 3 hours 1 minute, respectively. The latter time intervals were covered
with the LOFAR international stations (see below).

At GMRT frequencies, \psr\ is a strong radio source and the main radio pulse can be used to readily identify mode switches.
Very similar to the classification used by \citet{sobey2015}, we defined in this work the following radio modes, :

1) B mode: in this mode the pulsar is bright and displays short (a few pulses) nulls (cessation of emission). The main pulse (MP),
the pre-cursor (PC) and the inter-pulse (IP) can be easily seen, though the IP at 325 MHz is weak.

2) Null mode: the pulsar nulls in this mode and the average phase distribution does not show pulsed emission. Some sporadic spiky single-pulse emission across the mode cannot be ruled out.

3) Q mode: in the Q mode the pulsar shows sporadic and spiky single pulse emission in the region around the MP, including short nulls. The PC and IP are not seen at 325 MHz in this mode. 
This is consistent with \citet{sobey2015}, who called this behaviour RRAT-like.

4) Q-bright mode: in the Q mode we detected the onset of the mode to be associated with a bright sequence where both MP and PC can be seen, and the mode gradually fades to the Q mode.
  

 \begin{figure} 
 \begin{center}
   \includegraphics[width=8.5cm]{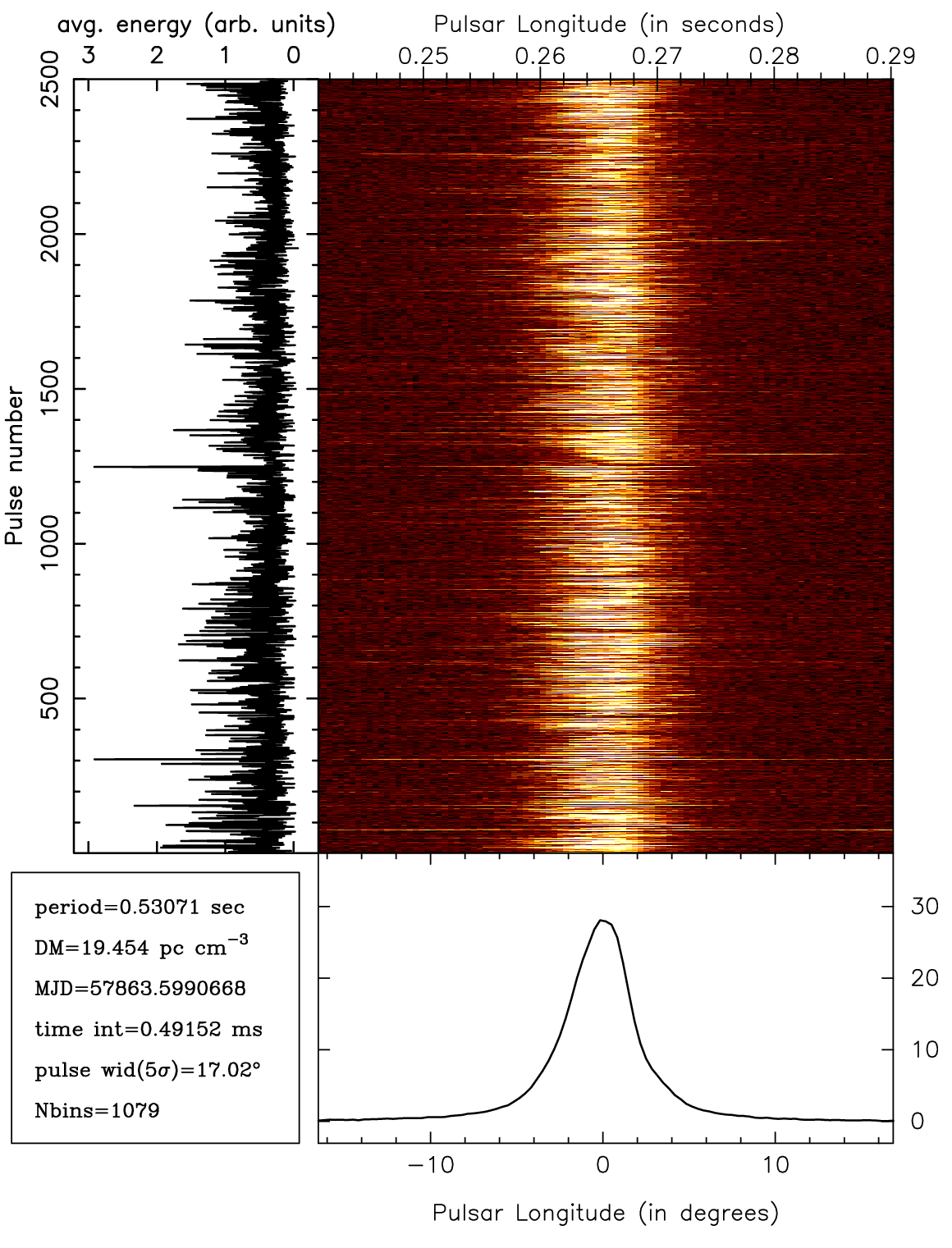}
   \caption{\label{April20GMRT306MHz} GMRT observation at 325 MHz of \psr\ on 2017 April 20,
   showing as a typical example \psr\ in B mode during 2500 single-pulse sequences, or $\sim$ 22 minutes of the total duration of $\sim$ 7.5 hours in B mode. 
   Observation time versus pulsar phase centred on the main pulse with underneath the
   integrated profile of the main pulse, and to the left the average energy per pulse in arbitrary units.  } 
  \end{center}
 \end{figure}
 
\begin{figure} 
 \begin{center}
   \includegraphics[width=8.5cm]{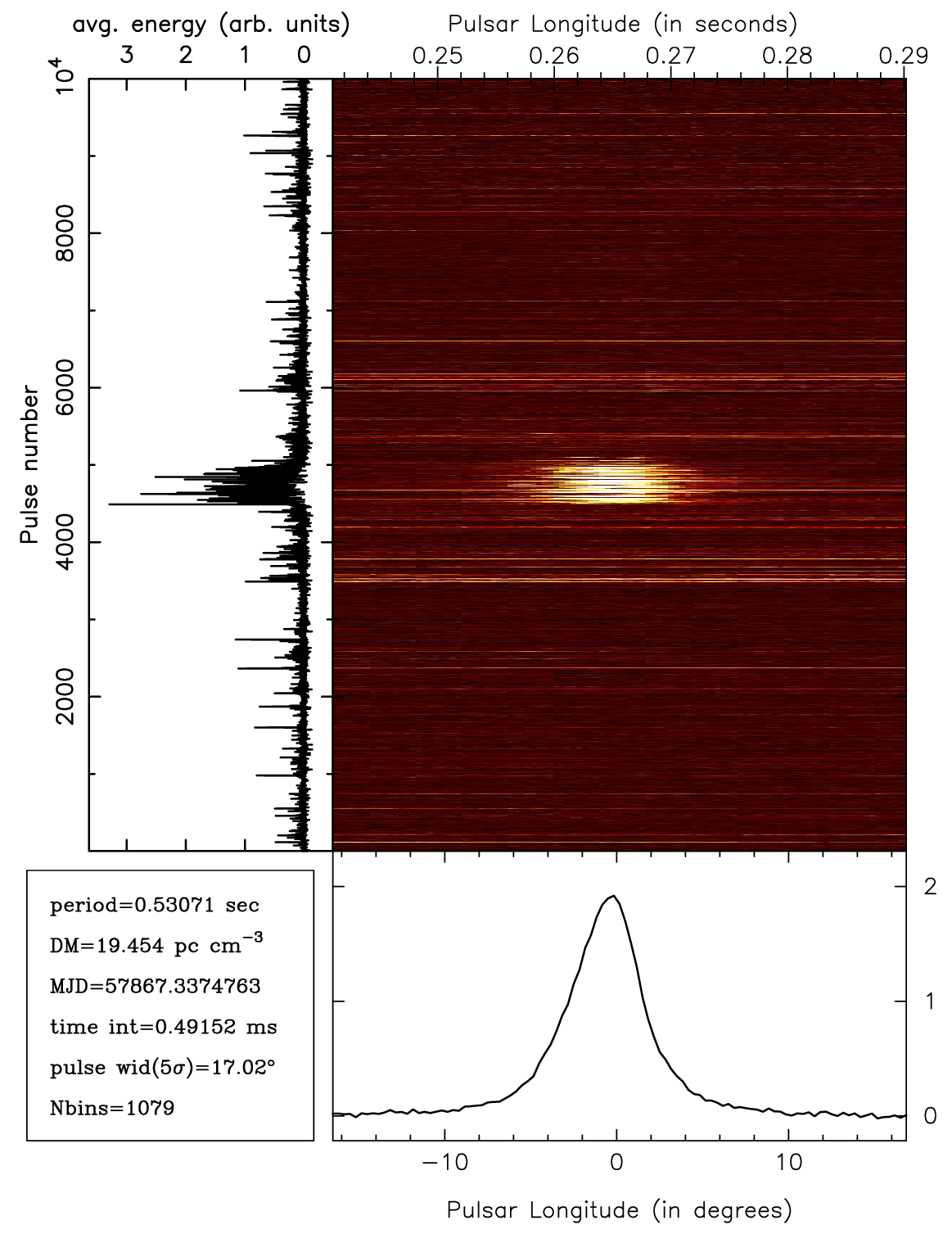}
   \caption{\label{April24GMRT306MHz} GMRT observation at 325 MHz of \psr\ on 2017 April 24,
   showing  the first 10,000 single-pulse sequences, or $\sim$ 88 minutes of the total $\sim$ 7.5 hours of observation time. 
   For explanation of figures, see caption of Fig.~\ref{April20GMRT306MHz}. The pulsar starts in the Null mode at pulse number 0, 
   switches into the Q-bright mode around pulse number 4500 and decays into the Q mode for the rest of the observation around pulse number 5250.
   Note: the energy scale of the pulses is arbitrary, and in this figure almost an order of magnitude smaller than that in Fig.~\ref{April20GMRT306MHz}. 
   An almost identical event can be seen in Fig. 3 of \citet{sobey2015}. } 
  \end{center}
 \end{figure}

Fig.~\ref{April20GMRT306MHz} shows an example of \psr\ detected in B mode. It shows 22 minutes of single-pulse sequences of the MP measured with the GMRT at 325 MHz in our first observation.
Unexpectedly, during five of our GMRT observations lasting on average $\sim$7.5 h \psr\ appeared to be continuously in the B mode (see Table~\ref{table_radio_obs}). Only in the observation on 2017 April 24 was \psr\
predominantly in the Q mode. In fact, the observations started when \psr\ was in the Null mode, then entered briefly the Q-bright mode and faded into the Q mode for the rest of the observation. 
These transitions are shown for the MP in Fig.~\ref{April24GMRT306MHz}.
The Q-bright mode lasted only $\sim$6-7 minutes, and the average pulse energy is lower by a factor 7-10 than the average measured in the B mode.
 

\subsection{LOFAR} 

For our campaign we were assigned 6 $\times$ 9 hours of observations at 150 MHz with the LOFAR \citep{stappers2011, vanhaarlem2013} DE601, FR606, SE607 and UK608 international stations in stand-alone mode (as part of project LC7-014), partly simultaneous with \xmm.
The stations provided redundancy and backup in the case of problems, also with the GMRT observations, and co-adding them incoherently can boost the S/N by a factor of $\sim$2.
It was only possible for the LOFAR observations to be simultaneous with \xmm\ and GMRT for the last hours of each of the \xmm\ observations. Table~\ref{table_radio_obs} lists the observation times with 
the LOFAR DE601 international telescope, representative for the observations with all participating LOFAR international stations. During all simultaneous GMRT and LOFAR observations there was agreement about the assigned radio mode of \psr. On April 22 and 30, the 
GMRT observations ended before the end of the \xmm\ observations by $\sim$3 hours. In both cases LOFAR showed that \psr\ continued in the B mode
for the remainder of the \xmm\  observations. In Table~\ref{table_radio_obs} one can see that LOFAR measured two mode transitions, Q to B mode on April 22, and B to Q mode on April 28. 2017.
Unfortunately, these transitions were not covered by \xmm.

\begin{table}
\caption{ {\it XMM-Newton} observation times of our six observations of \psr\ in 2017 April with the Pn and MOS cameras. Observation IDs 078440-0701, -0801, -0901, -1001, -1101, -1201, from top to bottom. The last column lists the 
dead-time-corrected exposures (lifetimes) per observation and for the total. The values listed for MOS apply to MOS 1 and MOS 2.
 }
\label{table_xmm_obs}
\begin{center}
\begin{tabular}{c c c c}
\hline
Date & Start Pn/MOS& End Pn/MOS &  Exp. Pn/MOS\\
yyyy-mm-dd  & hh:mm (UT)        &    hh:mm (UT)  & [ks]     \\
\hline
 2017-04-20 & 07:28 / 06:59 &   16:07 / 16:06 &   28.4 /  31.2\\
 2017-04-22 & 07:22 / 06:54  &  14:38 / 14:38 &  23.6 / 24.0 \\
2017-04-24 & 06:54 / 06:26 &  14:10 / 14:10 &   23.8 / 26.4 \\
2017-04-26 & 06:43 / 06:15 &  16:14 / 16:14  & 31.3 / 34.2  \\
2017-04-28 & 08:38 / 08:10 &  15:54 / 15:53 & 23.8 / 26.5 \\
2017-04-30 & 08:44 / 08:16 & 16:00 / 16:00   & 23.8 / 26.6 \\
Total &   &   &   154.7 / 168.9 \\
\hline \\
\end{tabular}
\end{center}
\end{table}


\section{{\it XMM-Newton} X-ray observations}

In April 2017, we obtained six \xmm\ observations of \psr\ with durations between
$\sim$7 and 9.5 h (see Table~\ref{table_xmm_obs}). 
We used only data obtained with the EPIC instrument 
aboard \xmm, which consists of one camera based on Pn CCDs \citep{struder2001}, the Pn, and two cameras based on 
MOS CCDs \citep{turner2001}, MOS~1 and MOS~2. The Pn camera was operated in Large Window mode, 
which provides a time resolution of 47.7 ms, and the MOS cameras in Small Window mode with a time resolution of 300 ms. 
The latter time resolution is insufficient for timing studies of \psr\ with its period of 530 ms.
For the three cameras we used the thin optical filter. We verified that our observations were not affected by increased 
particle backgrounds due to soft proton flares. The total lifetime (dead time corrected exposure) for the Pn
and MOS  cameras are also given in Table~\ref{table_xmm_obs}. The radio observations covered $\sim$ 87\% of the total \xmm\ observations.

\section{X-ray spatial analysis and mode switching}
\label{spatial}

In the spatial analysis, we apply a two-dimensional Maximum Likelihood (ML) 
method to the EPIC Pn and MOS 1 and 2 data using our knowledge of the two-dimensional point-source signature (the point spread function; PSF), and taking into account the 
Poissonian nature of the counting process. At first the events\footnote{Each event $i$ is characterised by its (barycentered) arrival time $t_i$, spatial coordinates 
$x_i, y_i$ , energy $E_i$, event pattern $\xi_i$ and flag $\digamma_i$. We used $\xi_i$ = [0,4] and flag $\digamma_i$ = 0 for both Pn and MOS.} are sorted
in two-dimensional count skymaps. The PSF is fitted to this count distribution at the known position of the pulsar
on top of a background structure, assumed to be flat. Typical fit-region sizes are of the order of 30-60$\arcsec$. For further details of the ML analysis that we applied 
to the EPIC Pn and MOS 1 and 2 data (treating these separately), see \citet{hermsen2017}.

\begin{figure}
  \begin{center}
     \includegraphics[width=8.5cm,height=14cm]{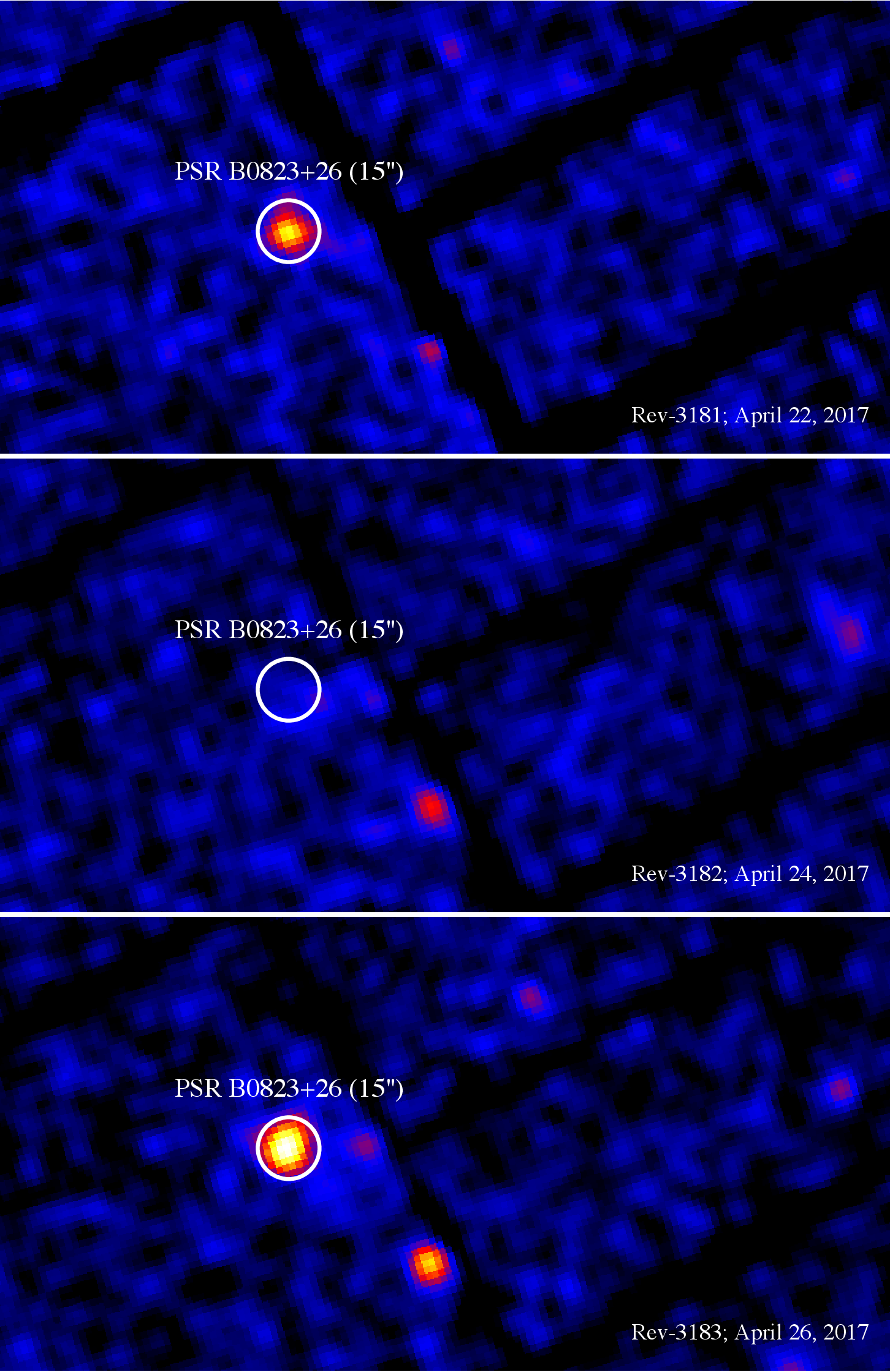}
     \caption{\label{EPIC Pn raw maps} \xmm\ EPIC Pn raw detector images (dimensions $\sim$ 7 by 4 arcmin) in celestial coordinates ($\alpha_{2000}, \delta_{2000}$) for energies 0.5$-$2 keV for the observations of \psr\ on 2017 April 22, 24 and 26
     (top to bottom, \xmm\ revolutions 3181, 3182, 3183, respectively). The white circles with radii 15 arcsec are centred on the radio position of \psr. On April 22 and 26 when the
     pulsar is in the radio B mode, \psr\  is evidently present. On the contrary, on April 24 \psr\ is in the radio Q mode and is not visible in the X-ray map for a similar exposure. 
     The orientation is: North is up and East is left.}
  \end{center}
\end{figure}

Given the detection with \xmm\ of \psr\ by \citet{becker2004} and assuming that \psr\ behaves as a stable emitter in X-rays, we expected to detect the pulsar in each of the six separate observations in the raw detector images. 
In the first step of the analysis we produced for each observation such raw detector images for energies 0.5$-$2 keV, and surprisingly already obtained 
convincing evidence of synchronous X-ray/radio mode switching of \psr. 
In the maps of the five observations during which the pulsar was shown to be
in the radio-B mode, indeed, the pulsar was clearly visible, however, for the radio-Q-mode observation on April 24 
the source was not visible in the raw images. 
 Fig.~\ref{EPIC Pn raw maps} shows as an example the
raw images for the EPIC Pn detector, energies 0.5$-$2 keV, for the observations on April 22, 24 and 26. 
Application of the ML analysis to the calibrated data of the radio B-mode observations, including energies down to 0.2 keV, rendered detection significances (energies 0.2$-$2 keV) ranging from 11.7$\sigma$ to 19.2$\sigma$.
For the observation on April 24, including the short Q-bright mode, the ML analysis showed a much weaker 3.0$\sigma$ detection for a similar exposure. 

 \begin{figure*} 
 \begin{center}
   \includegraphics[width=7.5cm,height=14cm,angle=90]{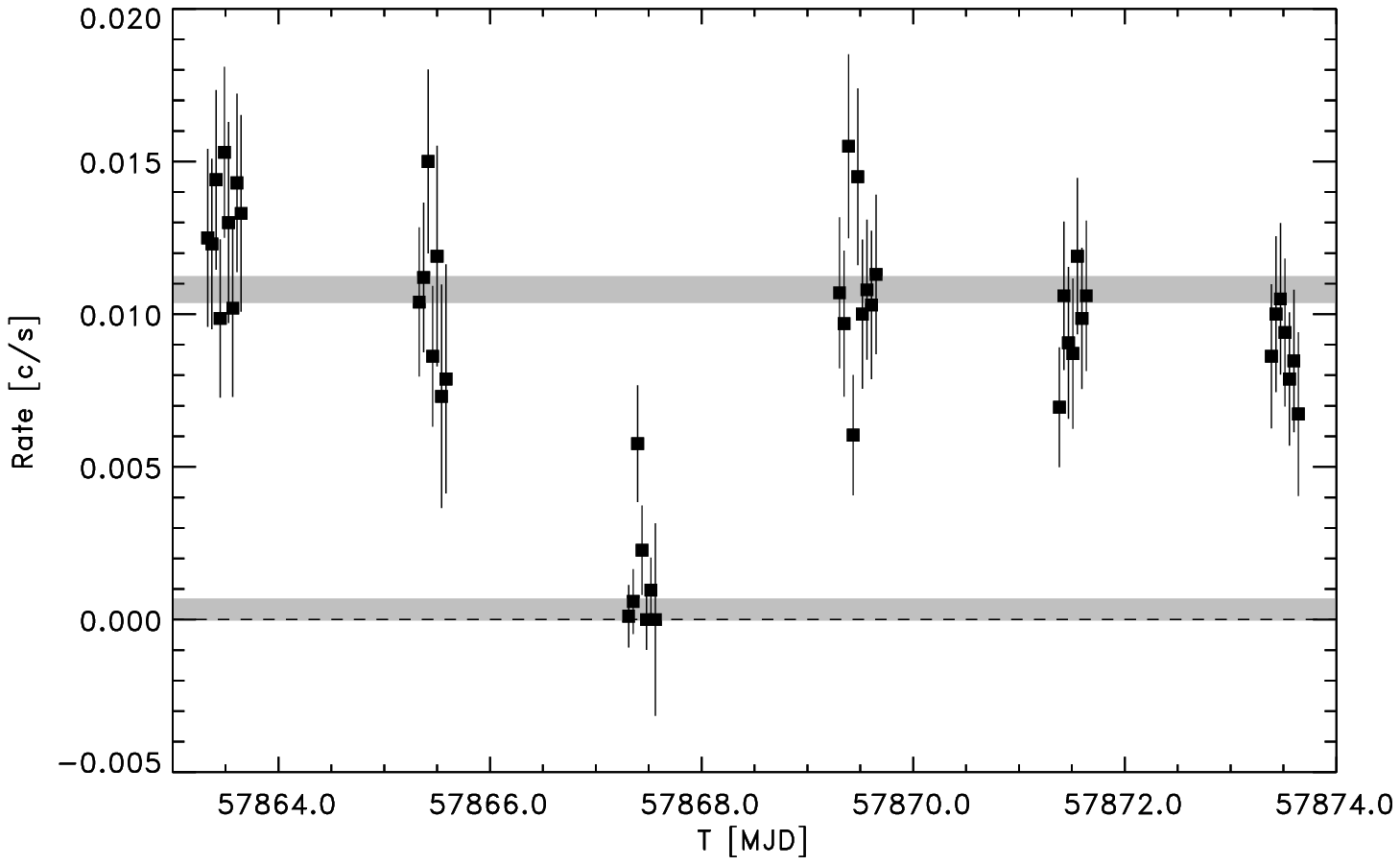}
   \caption{\label{PSRB0823lightcurve} \psr\ count rates (0.2$-$2 keV) in time intervals of $\sim$1 hour as a function of Modified Julian Day for all our \xmm\ EPIC Pn observations from 2017 April 20 to April 30.  Error bars are $\pm 1\sigma$.
  The upper gray band shows the average count rate during the B-mode observations, ($10.81 \pm 0.44) \times 10^{-3}$ counts s$^{-1}$, and the lower band the average count rate ($0.38 \pm 0.42) \times 10^{-3}$ counts s$^{-1}$, thus consistent with no detection) during the Q-mode observation on April 24 excluding the only 1-hour interval in which \psr\ was significantly (4.6$\sigma$) detected.
  The band widths correspond to 2$\sigma$.} 
  \end{center}
 \end{figure*}
  


 \begin{figure} 
 \begin{center}
   \includegraphics[width=8cm,height=7cm]{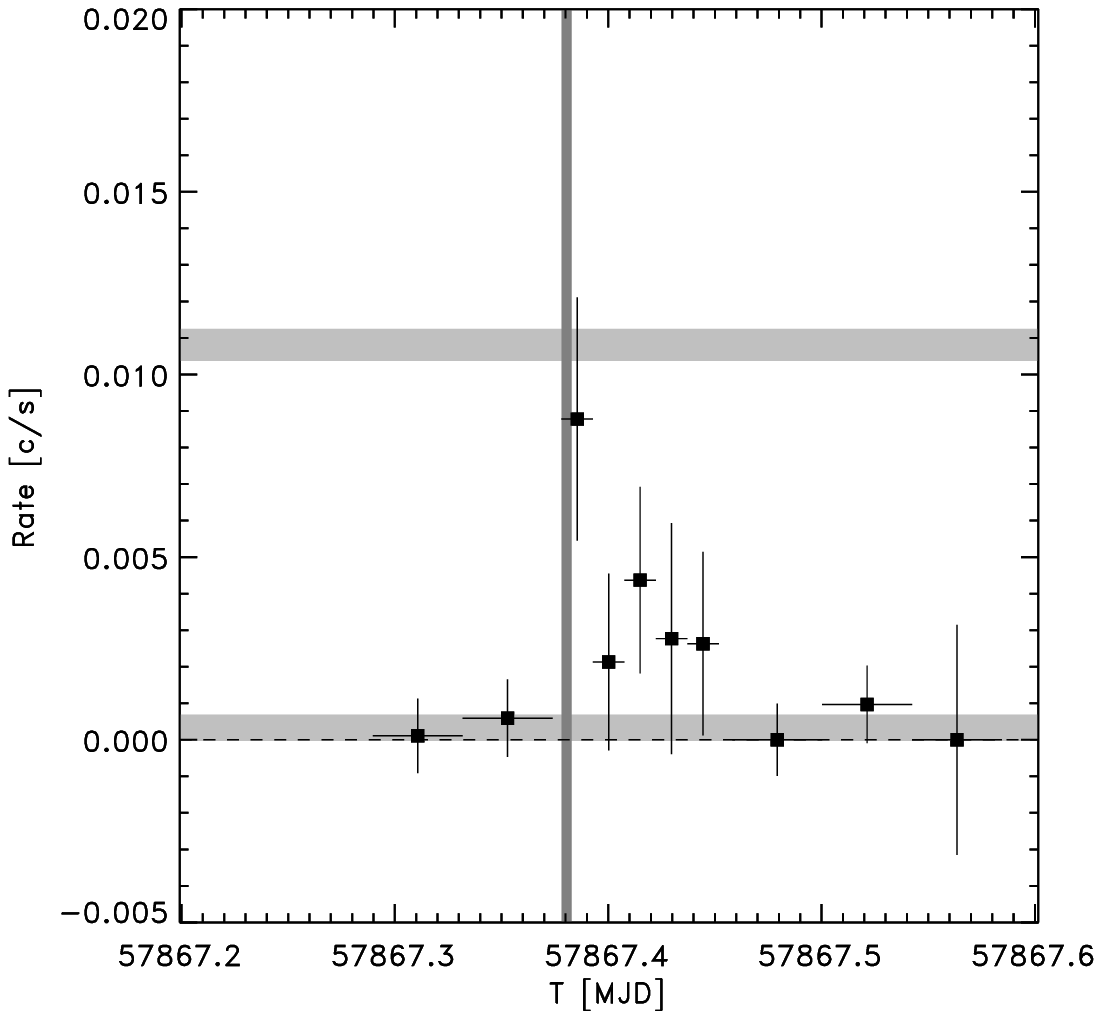}
   \caption{\label{April24ZOOM} \psr\ count rates (0.2$-$2 keV; EPIC Pn) for our observation of 2017 April 24, when \psr\ was predominantly in the radio Q mode. The first two X-ray values are measured during
   the initial radio null mode (no significant detection radio and X-ray emission),
   followed by a short 6-7 minutes interval with enhanced radio emission (Q-bright mode, see Fig.~\ref{April24GMRT306MHz}), before \psr\ entered into 
   the Q mode. The vertical gray band indicates the onset and six minutes of the Q-bright mode. The five time bins starting at the onset of the Q-bright mode are 21 minutes wide. The other time bins are 1-hour wide.   
  Error bars are $\pm 1\sigma$. For the horizontal grey bands, see Fig.~\ref{PSRB0823lightcurve}. } 
  \end{center}
 \end{figure}
  


 \begin{figure} 
 \begin{center}
   \includegraphics[width=8cm,height=7cm,angle=0]{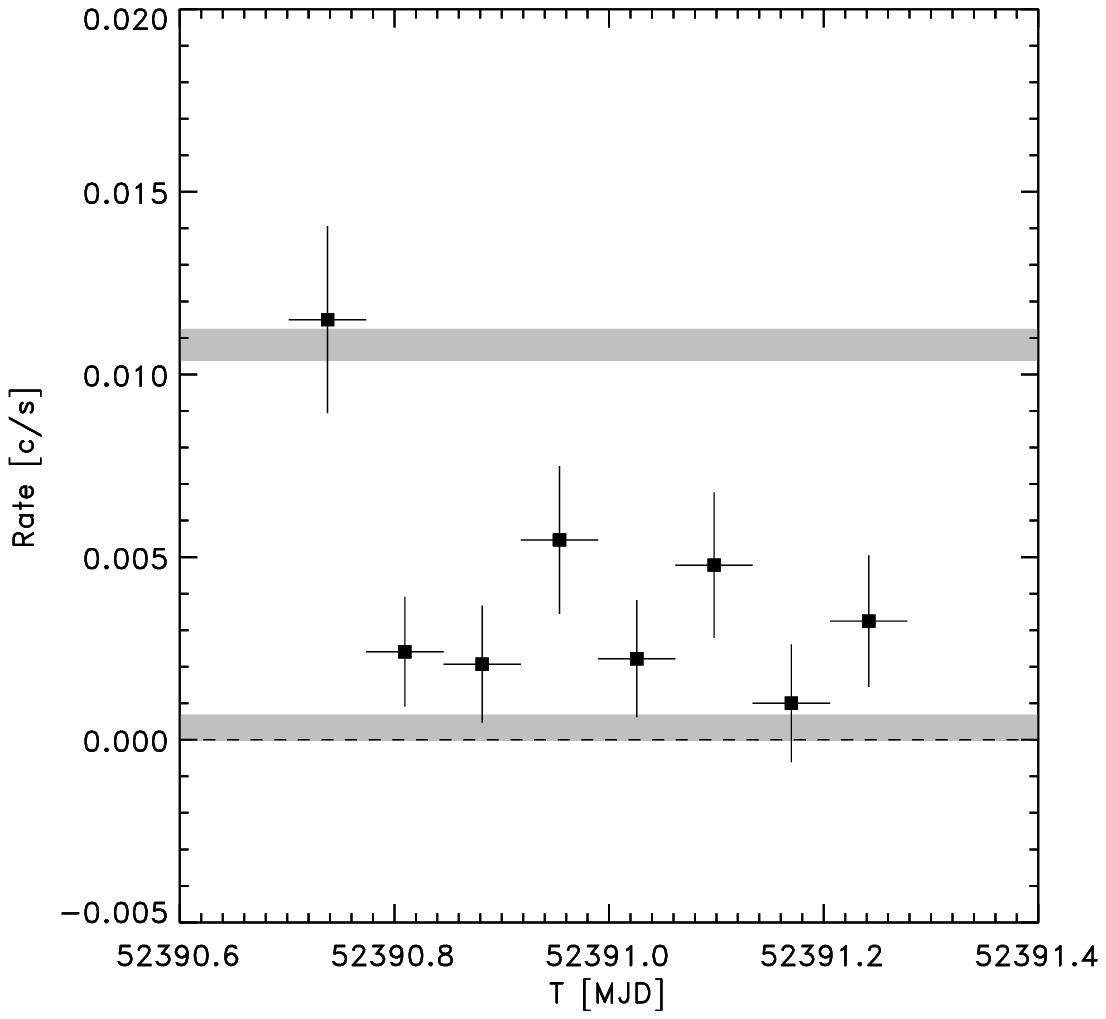}
   \caption{\label{Beckerlightcurve} \psr\ Pn count rates (0.2$-$2 keV) for the archival observation of 2002 April 26 \citep{becker2004}. Error bars are $\pm 1\sigma$.
  The gray bands are the same as in Fig.~\ref{PSRB0823lightcurve}, and show the average count rates measured during our B-mode observations of 2017 April, and during our Q-mode observation on 2017 April 24, respectively. The band widths
  correspond to 2$\sigma$.}
  \end{center}
 \end{figure}
  

In oder to investigate this further, we show in Fig.~\ref{PSRB0823lightcurve} the pulsar count rates (0.2$-$2 keV) in time intervals of $\sim$1 hour for all our observations. 
It is evident that the X-ray count rate 
gives a direct indication of the actual radio B or Q mode of \psr. Furthermore, Fig.~\ref{PSRB0823lightcurve} shows that the X-ray count rates at the start of each observation where we did not have radio coverage, are consistent with
the values measured during the rest of the observations, whether \psr\ is then in the B mode or the Null or Q mode. Therefore, we can conclude that \psr\ was at the start of the observations on April 20, 22, 26, 28 and 30
already in the radio B mode, and on April 24 in the radio Null or Q mode. We have adopted this classification in the further X-ray analysis when selecting X-ray events in observation windows with \psr\ in the radio B or Q mode.

One can see in Fig.~\ref{PSRB0823lightcurve} that during the Q-mode observation six of the seven flux values are within 1$-$2$\sigma$ consistent with no detection. Only one value stands out significantly. Selecting the data collected in the corresponding
$\sim$1-hour time interval and applying the ML analysis gives a 4.6$\sigma$ source detection. Apparently, the 3$\sigma$ detection in the total observation of April 24 was due to \psr\ being active in this 1-hour interval.
The upper gray horizontal band shows the average count rate during all observations with \psr\ in the B-mode ($10.81 \pm 0.44) \times 10^{-3}$ cnts/s. The lower horizontal band 
shows for the Q-mode observation of April 24 the average count rate during that observation but excluding the 1-hour interval in which \psr\ was active ($0.38 \pm 0.42) \times 10^{-3}$ cnts/s, consistent with no detection or 2$\sigma$ upper limit of
$1.22 \times 10^{-3}$ cnts/s).

It is interesting to note in Fig.~\ref{PSRB0823lightcurve} that the count rate (0.2 $-$ 2 keV) over the five days of our campaign with \psr\  in the B mode does not seem to be constant. 
The average count rate measured during the first observation on April 20, ($13.0 \pm 1.0) \times 10^{-3}$ cnts/s, is higher by $\sim 3.0 \sigma$ than that during the last day 
on April 30, ($9.0 \pm 0.9) \times 10^{-3}$ cnts/s. If we approximate the evolution of the count rate by a linear decay, the best fit gives an improvement by 3.1$\sigma$
compared to a constant count rate. It is, however, more likely that \psr\ exhibits, during its radio B mode, some variability in its X-ray flux of order $\pm$ 20\% on time-scales of hours or days.

In Fig.~\ref{April24ZOOM} we reproduce the light curve for the Q-mode observation on April 24, 2017, with shorter time intervals of 21 minutes when our counting statistics allowed this. 
\psr\ exhibited a radio null mode at the start of the observation until the onset of the 
short 6-7 minutes Q-bright mode (vertical gray band in Fig.~\ref{April24ZOOM}), followed by the Q mode for the remaining $\sim$ 4.5 hours. 
In this figure we selected the shorter time intervals starting at the onset of the radio Q-bright mode.  
In the first of the short 21-minutes intervals (MJD 57867.38) we detected \psr\ in the X-ray sky map at a significance level of 4.5$\sigma$. We did not detect
\psr\ in X-rays during the preceding radio null mode, thus after the onset of the short radio Q-bright mode we measured significant correlated flaring in X-rays with a possible decay over $\sim$1.5 hour
to measured count rates consistent with zero. 
 
 In the Introduction we noted that \citet{becker2004} detected \psr, but with an X-ray count rate that appears now to be significantly lower than the count rate we measure in the radio B mode,
and somewhat higher than we measure in the Q mode. To investigate any (in)consistency, we revisited the archival \xmm\ observation of April 26, 2002, and produced the light curve shown in Fig.~\ref{Beckerlightcurve}. 
For easy comparison, we show the average count rates of
the B and Q modes as measured in our campaign and drawn in Fig.~\ref{PSRB0823lightcurve}. From this comparison, we conclude that \psr\ was in the radio B mode at the start of the observation in 
April 2002 with the same count rate as measured on average in the B mode during our campaign.
During the remainder of the 2002 observation, the measured count rates are similar to the values we measured in 2017 April 24 in 
the Q mode during the decay after the high count rate in the Q-bright mode (see Fig.~\ref{April24ZOOM}),  but most individual values do not represent significant detections.

\begin{table}
\caption{Jodrell Bank ephemeris of \psr\ valid for our \xmm\  observations.}
\label{ephemeris}
\begin{center}
\begin{tabular}{c c}
\hline
Right Ascension (J2000)   & $ 08^{\hbox{\scriptsize h}} 26^{\hbox{\scriptsize m}} 51\fs538 $ \\
Declination (J2000)    &   $ +26\degr 37\arcmin 21\farcs35$   \\
Epoch (TDB)    & 57579  \\
$\nu $~(Hz)& 1.8844381708762   \\
$\dot{\nu}$~(Hz~s$^{-1}) $  & $-5.96902 \times  10^{-15}$\\
$\ddot{\nu}$~(Hz~s$^{-2}) $  & $-1.95 \times 10^{-24}$  \\
 Start (MJD) & 57306 \\
 End (MJD) &  57851\\
 Solar system ephemeris  &  DE405 \\
\hline \\
\end{tabular}
\end{center}
\end{table}

\section{X-ray timing analysis}
\label{timing}

For the timing analysis we could only use the Pn data, which had a sufficiently good time resolution of 47.7 ms, and selected events
detected within a $20^{\prime\prime}$ aperture around the X-ray position of \psr .
The times of arrival were converted to arrival times at the Solar System Barycentre and folded with the ephemeris of \psr\  given in 
Table~\ref{ephemeris}. We searched for a timing signature first in the individual observations, and discovered already in the first observation on
April 20, 2017, pulsed X-ray emission from \psr\  with the detection 
of a broad pulse in the energy band 0.2-2 keV at a significance of 9.2$\sigma$ \citep[$Z_1^2$ value,][]{buccheri1983}. The pulsed signal was detected in
all five B-mode observations, but not in the Q-mode observation on April 24, 2017. The latter is unsurprising, given that no point source was found
in the sky map of the Q-mode observation at the position of \psr. The 2$\sigma$ flux upper limit for the energy band 0.5-2 keV, 
derived in the spatial analysis for the Q mode data (excluding the bright-Q phase), 
amounts $1.48 \times 10^{-15} $erg cm$^{-2}$ s$^{-1}$, assuming a double-BB spectral shape as used for the B-mode (see Subsection \ref{Spectrum total}).


 \begin{figure} 
 \begin{center}
   \includegraphics[width=8.5cm, height=7.5cm]{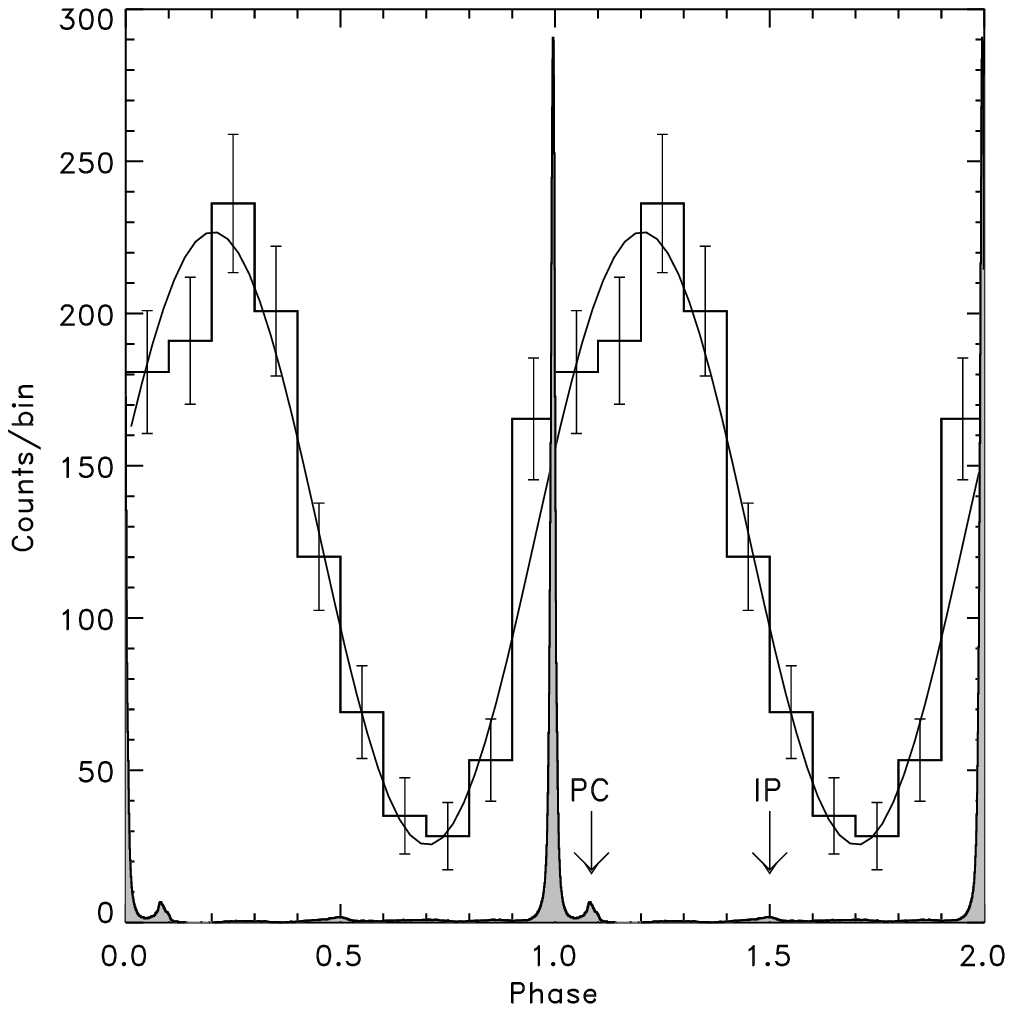}
   \caption{\label{ProfileB0823} X-ray pulse profile of \psr\  (histogram) using \xmm\ Pn data for all five B-mode observations for 
   the energy interval 0.2$-$2 keV, derived by phase-resolved imaging. The celestial background, assumed to be flat, is modelled out. 
   The y-axis gives in 10 phase bins the derived number of pulsed plus unpulsed counts from the point source.    Error bars are $\pm 1\sigma$. 
   The solid-line profile shows a fit with a sinusoid peaking at phase $0.203 \pm 0.012$. 
    For comparison, the radio profile, as measured with the GMRT at 325 MHz, is also shown. The strong radio main pulse peaks
at phase -0.005 and the phases of the much weaker Post Cursor (PC) and Inter pulse (IP) are indicated.    } 
  \end{center}
 \end{figure}
  


 \begin{figure} 
 \begin{center}
   \includegraphics[width=9.0cm,height=8cm]{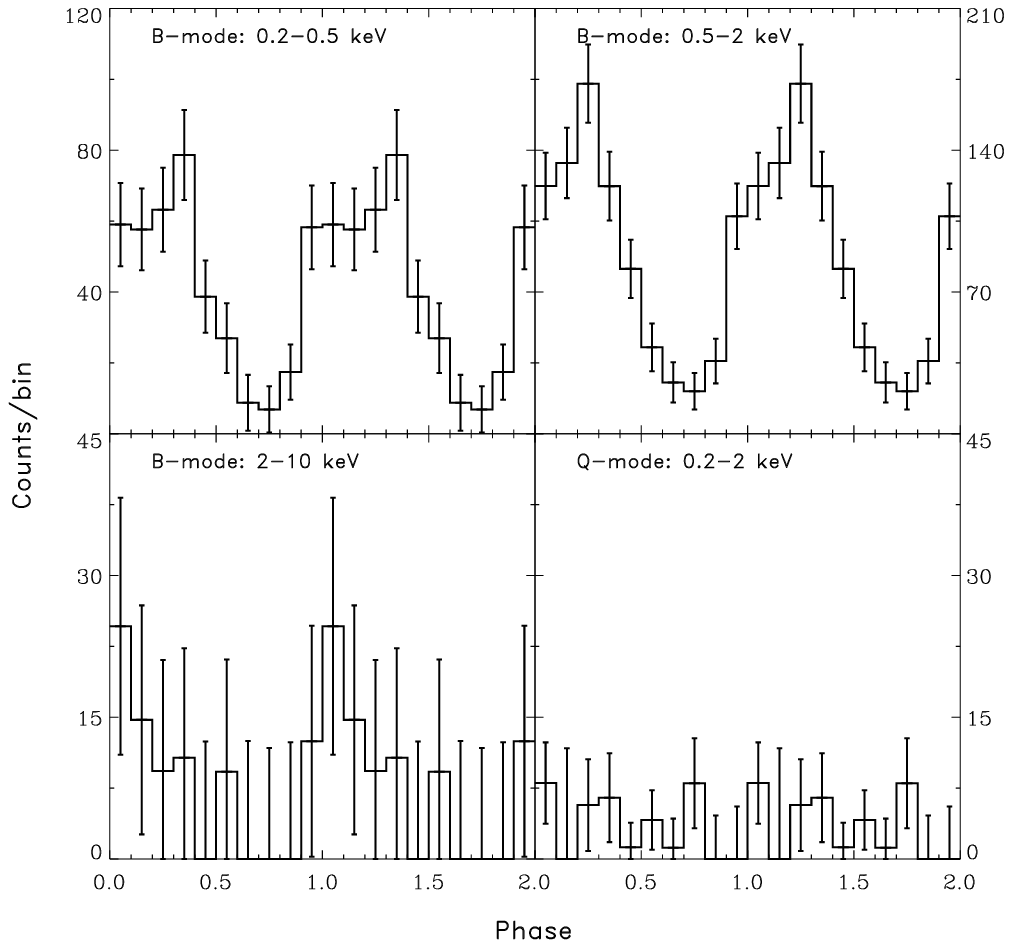}
   \caption{\label{ProfilesB0823} Pulse-phase distributions of \psr\ like in Fig.~\ref{ProfileB0823} combining all five B-mode observations  in differential energy intervals between 0.2 and 10 keV,   
   and for the single Q-mode observation of April 24 for energies 0.2$-$2 keV. A strong broad pulse is detected in
   the B mode between 0.2 and 2 keV, but no pulsation is found in the Q mode.
   Error bars are $\pm 1\sigma$. } 
  \end{center}
 \end{figure}
  

For the sum of all B-mode observations, Fig.~\ref{ProfileB0823} shows the X-ray pulse profile of \psr\ in the 0.2$-$2 keV energy interval. 
This pulse-phase distribution is generated by applying phase-resolved spatial analysis: 
for each phase bin count sky maps are produced and with the two-dimensional ML method the number of point-source counts is determined per phase bin
\citep{hermsen2017}. In this approach, the sky
background has been suppressed, and the phase distribution contains only source counts (pulsed plus unpulsed). 
The measured profile can be well fitted with a sinusoid and reaches its maximum at phase $0.203 \pm 0.012$. The radio profile, as we measured it with the GMRT at 325 MHz, with the strong main pulse
at phase $-0.005$ and the much weaker Post Cursor (PC) and Inter pulse (IP), is also indicated in Fig.~\ref{ProfileB0823}. The X-ray pulse lags the radio main pulse by $0.208 \pm 0.012$ in phase.

Fig.~\ref{ProfilesB0823} shows pulse profiles of \psr\ in three different energy intervals between 0.2 and 10 keV, and the phase distribution
for the Q-mode for energies 0.2$-$2 keV.  Fig.~\ref{ProfilesB0823} clearly shows for the B mode the detection of
the broad X-ray pulse for energies up to 2 keV, and no detection above 2 keV. In the Q mode (total observation on
April 24) there is, also in this representation, no indication for an X-ray pulse. The number of source counts detected in the
Q-bright interval is too low to construct a pulse profile.

\begin{figure} 
 \begin{center}
   \includegraphics[width=7.5cm, height=7cm]{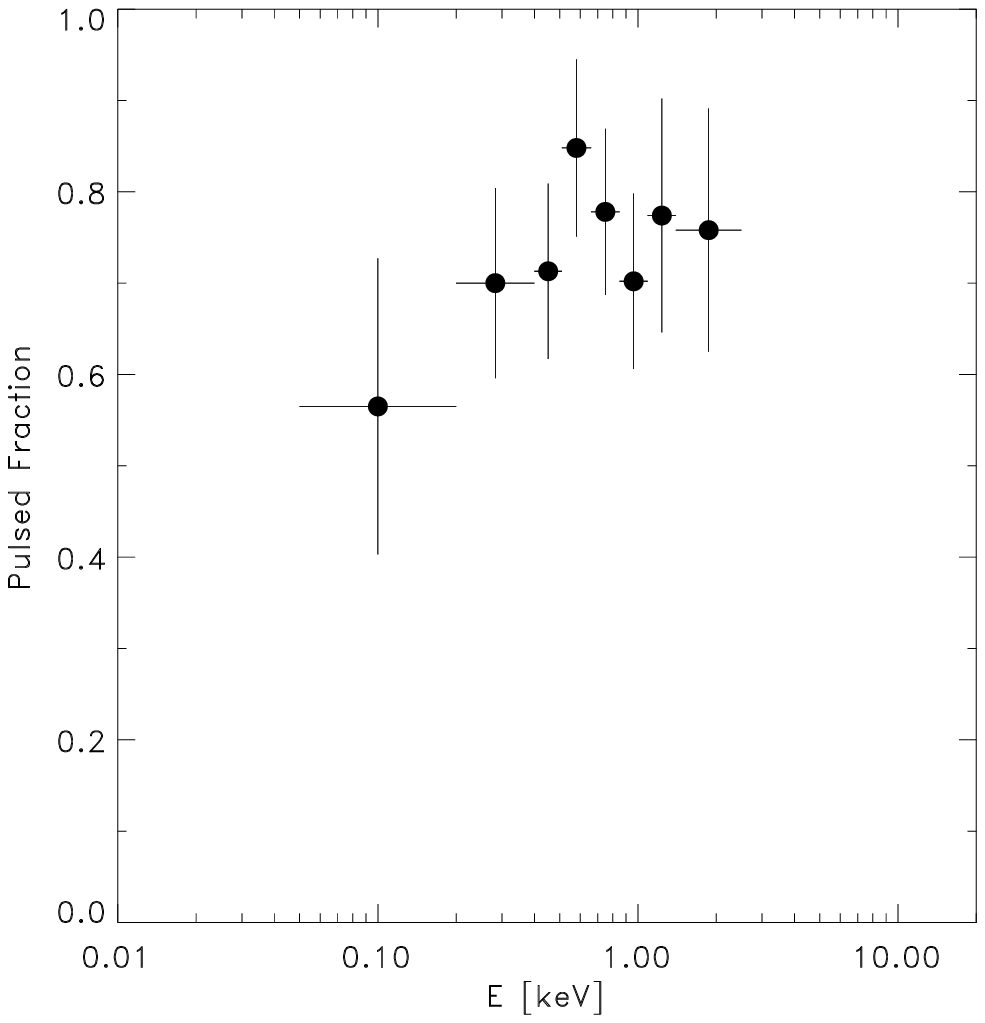}
   \caption{\label{B0823pulsedfraction} \psr\: pulsed fraction as a function of energy as derived from a three-dimensional ML analysis (see text).
   Error bars are $\pm 1\sigma$. } 
  \end{center}
 \end{figure}

From Fig.~\ref{ProfileB0823} it is evident that the measured X-ray pulsed fraction for energies 0.2$-$2 keV is very high. To derive the pulsed fraction as a function of energy, we applied a three-dimensional ML method:
assuming that the pulse profile is sinusoidal across the \xmm\ energy band, we can generalize the two-dimensional ML method by also taking into account the pulse-phase information of the events in a three-dimensional approach
(see for details \citet{hermsen2017}).
The derived pulsed fraction as a function of energy is shown in Fig.~\ref{B0823pulsedfraction}. The values between 0.2 and 2 keV vary
over the range $\sim$70 to 80\%.


 \begin{figure*} 
 \begin{center}
   \includegraphics[width=7cm,height=14cm,angle=90]{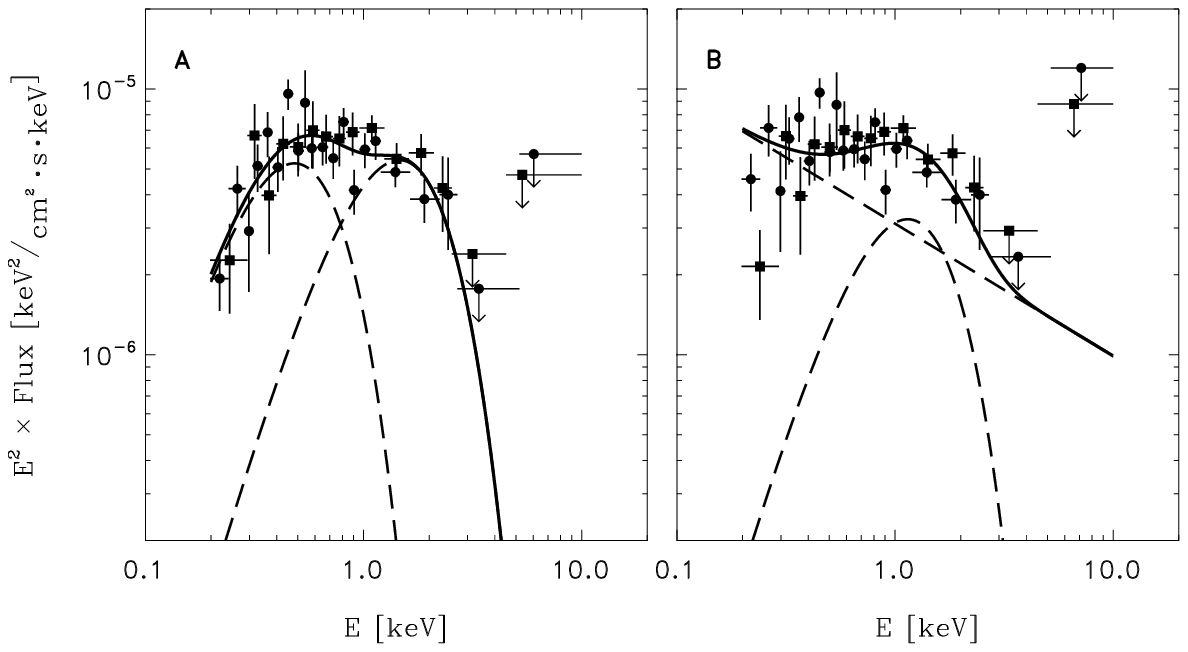}
   \vspace*{10mm}
   \caption{\label{totalspectra} Unabsorbed X-ray photon spectrum of the total emission of \psr\  in the radio B mode, derived from spatial analysis for
   \xmm\ EPIC Pn (filled circles) and MOS 1 and 2 (filled squares). {\it Panel A}: The solid line shows the best fit with two blackbody components ($\sim$ 30\% goodness-of-fit probability); the broken lines show the contribution from a soft
   ({\it kT} = 0.12 keV) and a hard component ({\it kT} = 0.37 keV), respectively. {\it Panel B}: Similarly, best fit ($\sim$ 10\% probability) with a blackbody ({\it kT}=0.29 keV) and a power law (index=$-$2.5).
   Error bars are $\pm$ 1$\sigma$.}
  \end{center}
 \end{figure*}
  

\section{X-ray Spectral analysis}
\label{Spectral analysis}
We excluded from all our spectral analyses the observation on April 24, 2017, when the pulsar was successively in the Null, Q-bright and Q mode, and was not detected 
in the X-ray skymap of the total observation. The X-ray detection in the short Q-bright mode had too few source counts to derive a spectrum as well. Thus, we
concentrate on the B mode.

\subsection{Spectrum total emission of \psr }
\label{Spectrum total}

We first derived the spectral parameters for the total emission from \psr\ between 0.2 and 10 keV using all the observations in Table~\ref{table_xmm_obs} during
which \psr\ was in the B mode.  We used the Pn and MOS 1 and 2 data
and performed again the two-dimensional ML analysis on the counts sky maps for user-defined energy windows between 0.2 and 10 keV, again adopting $\xi=[0,4]$ and $\digamma=0$ (see footnote in Section \ref{spatial}).
EPIC Pn and MOS 1 and 2 data were treated separately. 
We fixed the absorbing column density, $ N_{\hbox{\scriptsize H}} $, towards \psr\ in all our spectral modelling to $3.43 \times 10^{20} $cm$^{-2}$  \citep{kalberla2005, dickey1990}.

We first produced fits to the total source-count spectrum with a single power law (PL; ${\alpha}E^{\Gamma}$, with $\alpha$ the normalisation and $\Gamma$ de index) and a 
single blackbody (BB). Both fits are not acceptable with $\chi_{r}^2$ = 1.91 (for 38 degrees of freedom (dof)) and $\chi_{r}^2$ = 2.97 (for 38 dof), respectively.
A fit with two blackbodies gave an excellent fit with $\chi_{r}^2$ = 1.12 (for 36 dof): in this, we use a cooler blackbody with $kT \sim 0.12$ keV with hot spot radius $R$
of $\sim 77$ m, and a hotter blackbody with $kT \sim 0.37$ keV with $R$ only $\sim 8.5$ m (see Table~\ref{spectral_fits}a and Fig.~\ref{totalspectra}a). 
However, a fit with a BB plus PL also gave a good fit with $\chi_{r}^2$ = 1.29 (for 36 dof): $kT \sim 0.29$ keV with $R$ 
 $\sim 11$ m and $\Gamma \sim -2.50$ (see Table~\ref{spectral_fits}b and Fig.~\ref{totalspectra}b).
The double-BB fit is preferred over the BB-plus-PL fit (goodness of fit probability $\sim$30\% versus  $\sim$10\%).

For estimating the spectral shape of the total spectrum, we could use our maximal statistics by using the events detected by the Pn and MOS 1 plus 2. For the spectra of components of the
pulsar-phase distribution, we need the timing information and can only use the Pn events. In order to see whether there are systematic differences between the derived parameters for fits to the total spectrum of \psr\
using Pn and MOS 1 plus 2 data compared to using just Pn data, we repeated the double-BB and BB-plus-PL fits for the latter case. Table~\ref{spectral_fits} shows that we get
``identical' fit parameters, and that again the double-BB fit ($\sim$5\% probability) is preferred over the BB-plus-PL fit ($\sim$1\% probability). In conclusion, we do not see a systematic difference in the fit results, but mainly 
a loss in counting statistics when only the Pn data is used.

\begin{table*}
\caption{ Spectral fits to the total 
emission spectrum and to the pulsed and unpulsed ($DC$) emissions of \psr\ in the B mode over the energy range 0.2-10 keV using
MOS~1\&2  and/or Pn data of all XMM-Newton observations of the 2017 campaign. For the pulsed
emission, two cases are considered: 1) the pulsed counts are defined as the excess counts in the main X-ray pulse above a flat background level, {\it Pulse-ME};
2) it is assumed that there are two sinusoidal pulses in the light curve, namely one strong pulse at the main X-ray pulse, {\it Pulse-M}, 
and one corresponding to an antipodal inter pulse, {\it Pulse-IP}. Table {\bf I}), fits with two black-body models (BB); Table {\bf II}), 
fits with a power-law (PL) and black-body (BB) model.
$N_H$ has been kept fixed at a value of $3.43 \times 10^{20} $cm$^{-2}$. Values for $kT$ in bold are kept fixed (see text).}
\label{spectral_fits}
\begin{center}

\begin{tabular}{cccccccc}
\hline
 & Model  & Total &  Total &  Pulsed & DC & Pulsed & Pulsed   \\
  & Fit par.  &  Pn/MOS1+2 &   Pn &  {\it Pulse-ME} &  & {\it Pulse-M}   &  {\it Pulse-IP}    \\
\hline
 {\bf I} &  &  &  &  & & & \\
 &   & BB+BB &   BB+BB   &    BB+BB  & BB+BB &  BB+BB & BB+BB  \\
& $\alpha_{BB_1}$$\times 10^2$&   $0.48 \pm 0.17$  &   $0.60 \pm 0.27$  &   $0.31 \pm 0.19$  & $0.14 \pm 0.05$ &    $0.39_{-0.20}^{+0.43}$ &  $ 0.07 \pm 0.02 $ \\
& $kT_1$   & $0.12 \pm 0.01$  &    $0.12 \pm 0.01$  &  $0.13 \pm 0.03$  &  {\bf 0.12}  &  $ 0.13 \pm 0.02 $ &  {\bf 0.12}  \\
& $R_{BB_1}$    & $77 \pm 14$  &   $ 86 \pm 20$ &   $ 62 \pm 19$ & $41.5 \pm 7.4$ & $70_{-18}^{+38}$  & $29.4 \pm 4.2$  \\
& $F_{BB_1}$ $\times 10^{15}$   & $4.5 \pm 0.5$ & $ 4.2 \pm 0.5$& $4.4 \pm 0.7$  &  $1.1 \pm 0.4 $ & $ 4.8 \pm 0.8$ & $ 0.6 \pm 0.2$ \\
& $\alpha_{BB_2}$ $\times 10^5$&   $5.90 \pm 2.25$  &   $6.54 \pm 3.50 $ &   $2.55 \pm 3.40$   & $ 1.3 \pm 0.5$ &  $3.0_{-1.81}^{+4.22}$ &  $ 0.65 \pm 0.25$  \\
& $kT_2$    &    $0.37 \pm 0.03$ &   $0.36 \pm 0.04$ &   $0.41 \pm 0.13 $    & {\bf 0.37}  & $0.41_{-0.09}^{+0.13}$ & {\bf 0.37} \\
& $R_{BB_2}$    & $ 8.5 \pm 1.6$ &   $ 8.9  \pm 2.4$ &  $ 5.6 \pm 3.9$ &  $4.0 \pm 0.8$ & $ 6.1_{-1.8}^{+4.3}$  & $2.8 \pm 0.5$ \\
& $F_{BB_2}$ $\times 10^{15}$         &    $8.5_{-0.9}^{+0.8}$  &  $ 8.0 \pm 1.0$        &  $4.9 \pm 1.2$   & $ 1.8 \pm 0.7$  &  $ 5.9 \pm 1.4$  &  $ 0.92 \pm 0.35$   \\
& $\chi_{r}^2$ / dof   &  1.12 / 36 &  1.60 / 17 &   0.30 / 4  & 0.54/6 &  0.34 / 4  & 0.54/6   \\
\hline
{\bf II} &   &  &  &  &  &  &  \\
& &  BB+PL &    BB+PL  &   BB+PL & BB+PL & BB+PL & BB+PL  \\
& $\alpha_{BB}$$\times 10^3$&  $0.09 \pm 0.04$   &   $0.13 \pm 0.09$    &   $0.4 \pm 0.7$ & $ 0.014 \pm 0.012$  &  $0.4 \pm 0.8$ &  $0.007 \pm 0.006$\\
& $kT$   &   $0.29 \pm 0.03$  & $0.25 \pm 0.05$   & $0.18 \pm 0.05$  &   {\bf 0.29}   &   $0.18 \pm 0.06$ &  {\bf 0.29} \\
& $R_{BB}$    &   $ 11 \pm 3$   & $13  \pm 5$   & $ 23 \pm 20$  &  $4.2 \pm 1.8$ &  $23 \pm 23$  & $2.9 \pm 1.2$ \\
& $F_{BB}$ $\times 10^{15}$   &      $5.1 \pm 0.9$    &  $ 3.6 \pm 1.0$    &  $3.0 \pm 1.0$ & $0.8 \pm 0.7$ & $2.8 \pm 1.1$  & $0.4 \pm 0.3$  \\
 &$\alpha_{Pl}$ $\times 10^6$&     $3.12 \pm 0.51$             & $3.44 \pm 0.67$      &  $2.66 \pm 0.75$ & $ 0.9 \pm 0.2$  &  $3.38 \pm 0.84$  & $0.5 \pm 0.1$\\
& $\Gamma$   &  $ -2.50 \pm 0.17$    &  $-2.39 \pm 0.20$      &   $-2.13 \pm 0.36$ & ${\bf -2.50}$ &   $-2.15 \pm 0.33$  & ${\bf -2.50}$ \\
 &$F_{Pl}$  $\times 10^{15}$  &   $7.5 \pm 0.7$  &  $8.2 \pm 0.8$     & $5.8 \pm 1.0$  & $2.0 \pm 0.5$  & $7.4 \pm 1.2$ &  $ 1.0 \pm 0.3$\\
& $\chi_{r}^2$ / dof   &   1.29 / 36 &   1.92 / 17   &   1.14 / 4 & 0.26/6  &  0.85 / 4  &  0.26/6\\
\hline
\multicolumn{8}{l}{Units: $N_H$ in cm$^{-2}$; $\alpha_{BB}$ in photons cm$^{-2}$ s$^{-1}$ keV$^{-3}$; $kT$ in keV; Fluxes $F$ are unabsorbed for the 0.5-2 keV band in erg/(cm$^2$s);} \\ 
\multicolumn{8}{l}{Hot spot radius $R_{BB}$ in meters adopting a source distance of 357 pc; $\alpha_{Pl}$ in photons cm$^{-2}$ s$^{-1}$ keV$^{-1}$ at 1 keV}\\
\end{tabular}
\end{center}
\end{table*}


\subsection{Spectra of the pulsed and unpulsed emissions of \psr }

\label{pulsed_unpulsed_spectra}

 In \citet{hermsen2017} we introduced the three-dimensional ML approach. In this approach the ML analysis 
is applied in the 3D data space where the axes are the two sky coordinates and pulsar rotational phase. The spatial PSF 
is the point source signature in the skymaps for differential energy intervals and e.g. a sinusoid the 
shape of  the pulse profile in all differential energy bins. 

However, in our spectral analysis we consider two possible interpretations/descriptions of the pulse-phase distribution, namely, the measured X-ray pulse profile (e.g. Fig.~\ref{ProfileB0823}) consists of two components: 1) a flat unpulsed
component at the level of the minimum in the profile (DC component), plus a sinusoidal pulse on top of the flat background level (Excess) at the phase of the X-ray Main pulse (ME);
2) two sinusoidal pulses, one at the position of the phase of the X-ray Main pulse (M), and one at the phase of the minimum of the pulse profile, which we call
the X-ray Interpulse (IP). The complementary DC and ME (case 1), or M and IP (case 2), source counts per differential energy interval are simultaneously determined in the three-dimensional ML approach, 
and in the next step converted to source flux values in the forward-folding spectral-fitting procedure. 

We first fitted for case 1, the combination of unpulsed emission DC with pulsed excess emission ME. The best fit assigns $\sim$20\% of the counts detected from \psr\ in the Pn data between 0.2 and 10 keV 
to the DC component and  $\sim$80\% to the ME pulsed emission. Both component spectra can be perfectly fitted with a single PL model with index ${\Gamma} = -2.2$. 
However, this is inconsistent with the spectrum of the total emission. Namely, the pulsed fraction
is close to 80\%, and we found that the total spectrum can not be explained with a single PL model, but that two components are required. Indeed, we can also obtain an excellent fit with a double-BB
model (see Table~\ref{spectral_fits}a) and with a BB-plus-PL model (see Table~\ref{spectral_fits}b) to the pulsed emission ME. The obtained temperatures and PL index are fully consistent with those derived for the 
total emission spectrum using our maximal statistics with Pn and MOS 1 plus 2 data. However, the number of free parameters went down from 36 to only 4, reflecting the loss in statistical accuracy due to the lower number 
of counts and energy bins available in the three-dimensional ML analysis. A spectral fit with two components to the very-low-statistics DC spectrum does not converge and we need to reduce the number of free parameters.
For comparison purposes, we fixed the two BB temperatures, or the temperature of a BB and the spectral index of a PL component, in fits to the DC spectrum to the values obtained for the total emission (see Table~\ref{spectral_fits}).

For case 2, fitting the spectra of an X-ray Main pulse (M) and an Interpulse (IP), together producing the measured pulse-phase distribution, we followed the same procedure as for case 1, 
with very similar results (see Tables~\ref{spectral_fits}). This is no surprise, because, by definition, the total counts assigned to M are $\sim$12.5\% higher than for ME, and the number
of counts assigned to IP are $\sim$ 50\% of the counts assigned to DC.

In general, the fits with two blackbodies (to the total spectra and the pulsed spectra) are better than with a BB plus PL. Also, the two temperatures and the radii of the hot spots derived in the fits to the total and pulsed spectra
are consistent, within 1$\sigma$. We conclude that the total spectrum and the pulsed emission are most likely the sum of two blackbodies, but an interpretation with a BB plus PL can not be excluded.

\subsection{Magnetized-atmosphere-model spectral fit to the total emission spectrum of {\psr}}

\label{atmosphere-model}


 \begin{figure} 
 \begin{center}
{\mbox{\includegraphics[width=7cm, height=7cm]{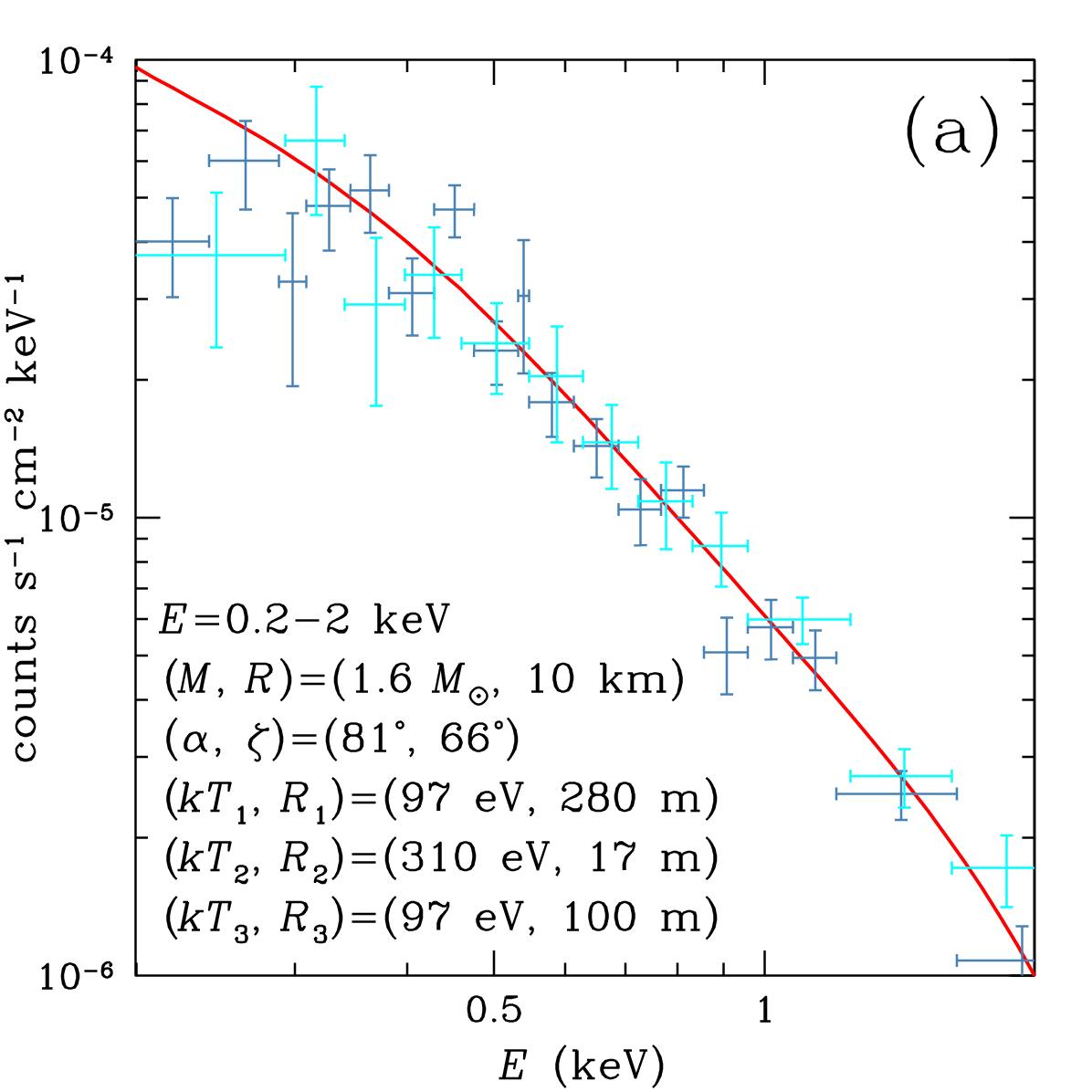}}}
{\mbox{\includegraphics[width=7cm, height=7cm]{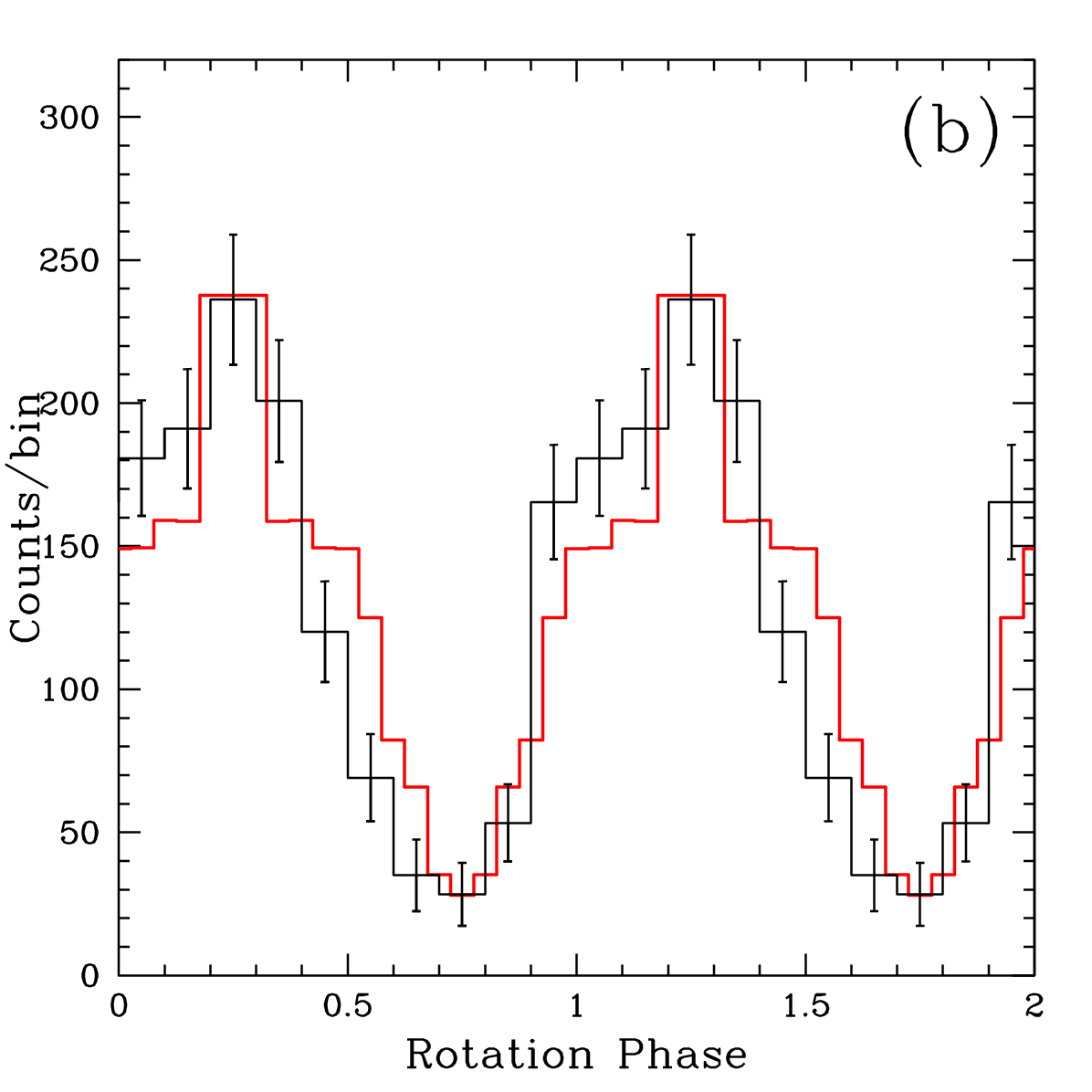}}}
{\mbox{\includegraphics[width=7cm, height=7cm]{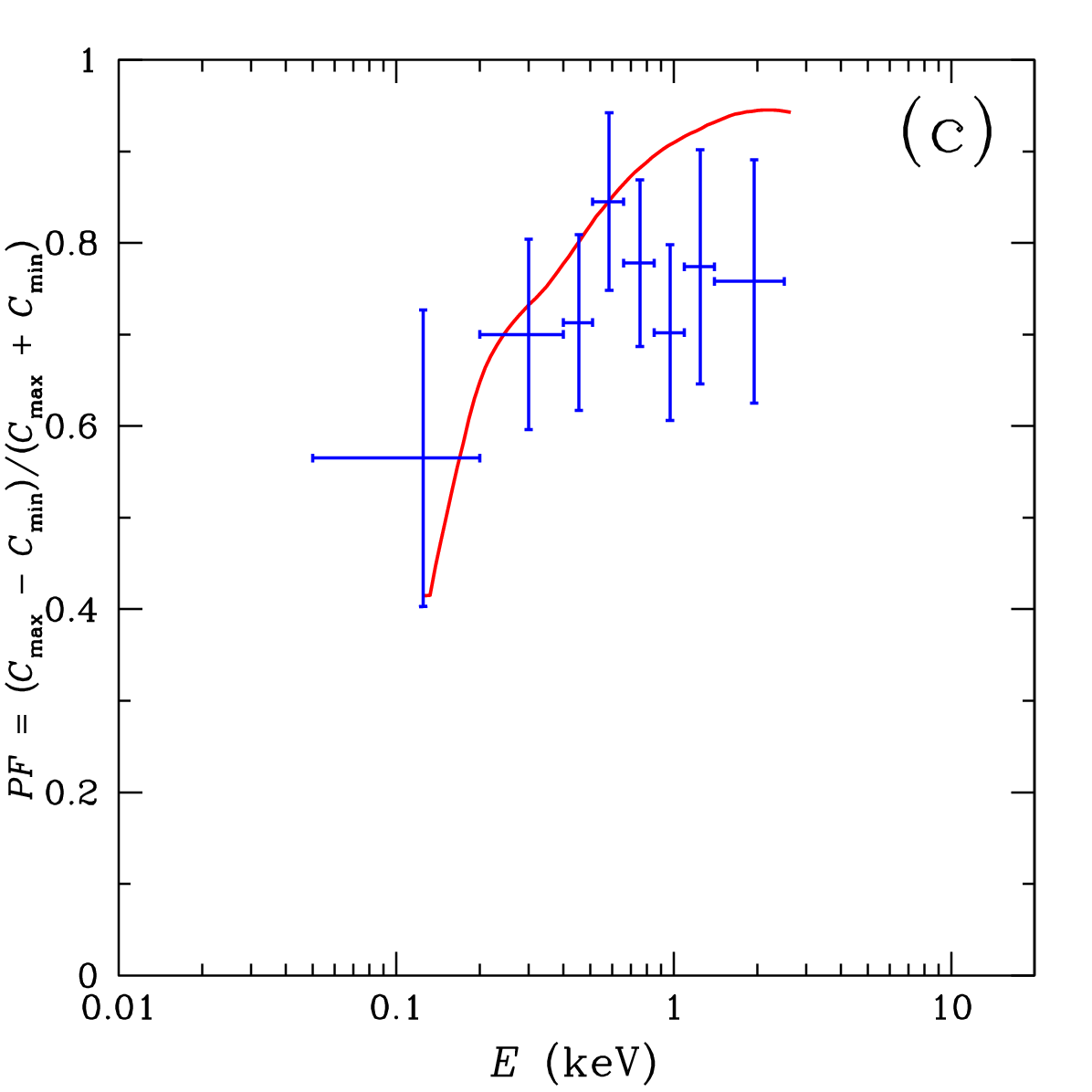}}}
 \caption{\label{spectr_atm} \psr\ in the radio B mode. a), unabsorbed X-ray photon spectrum of the total emission for
   \xmm\ EPIC Pn (dark blue) and MOS 1 and 2 (light blue); b), pulse profile for 0.2 $-$ 2 keV;  c), pulsed fraction $PF$. The red curves show the best fit with a (redshifted) magnetized partially ionized hydrogen atmosphere model. 
   For best fit parameters: see text and parameters in figure a). The definition of $PF$ in the model calculations (with $C_{max}$ and $C_{min}$ the maximum and minimum counts, respectively), 
   is for a sinusoidal pulse consistent with the definition used in the analysis. The model profile
   in figure b is approximately aligned with the observed one. Error bars are $\pm 1\sigma$. } 
  \end{center}
 \end{figure}
  

While the spectrum of PSR B0823+26 can be well-fit by the sum of two blackbody spectra (see
Subsection 6.2), it is more physically correct to describe thermal emission from
the surface of a hot neutron star by an atmosphere model.
Spectral fits using atmosphere models generally yield lower temperature and
larger emission radius than fits obtained using blackbodies
\citep{romani1987,shibanov1992}.
Furthermore, emission is anisotropic, with a beam pattern (i.e. specific
intensity as a function of photon propagation angle) that is energy-dependent,
especially when the atmospheric plasma is strongly magnetized, as in the case
of PSR B0823+26 with an inferred (equatorial) magnetic field of
$\sim 10^{12}\mbox{ G}$ (see Section 1).  This last point makes it possible
to achieve very high pulsed fractions, as we measured for {\psr}.
Therefore we employ a magnetized partially ionized hydrogen atmosphere model
\citep{ho2008,ho2014} to generate X-ray spectra and pulse profiles with a
procedure that accounts for relativistic effects \citep{ho2007}.
We consider a magnetic field of $10^{12}\mbox{ G}$ that is directed
parallel to the surface normal; field strengths of $5\times 10^{11}$ and
$2\times 10^{12}\mbox{ G}$ do not change our results qualitatively.
We assume a neutron star mass $M$ of $1.6M_\odot$ and radius of $10\mbox{ km}$,
so that the gravitational redshift is $1+z=1.38$.
We denote by $\alpha$ the angle between the magnetic and rotation axes and by
$\zeta$ the angle between the rotation axis and the line of sight, and
we note that $(\alpha,\zeta)$ and $(\zeta,\alpha)$ yield the same X-ray spectrum
and pulse profile.
For PSR B0823+26, fits to radio polarization data using the rotating vector
model give $(\alpha,\zeta)=(81^\circ,84^\circ)$, with an uncertainty of
$0.7^\circ$, for antipodal hot spots \citep{everett2001}.
However, models with these angles produce X-ray pulse profiles that have a
peak which is too narrow to match that obtained from \textit{XMM-Newton}
data (see Figure 7) because this geometry only permits the narrow component
(of width $<0.1$ in rotation phase; \citealt{pavlov1994}) of the beam
pattern to be visible.  Thus, while we assume $\alpha=81^\circ$, we allow
$\zeta$ to vary.

We are able to approximately match the X-ray spectrum, pulse profile, and pulsed fraction
of PSR B0823+26 with a model that includes three emission components:
a primary spot with a small hot core ($kT_2=310\mbox{ eV}$, $R_2=17\mbox{ m}$)
and a larger cooler annular concentric ring ($kT_1=97\mbox{ eV}$, $R_1=280\mbox{ m}$)
and an antipodal hot spot ($kT_3=97\mbox{ eV}$, $R_3=100\mbox{ m}$)
which largely produces the flux at pulse minimum (see Fig.~\ref{spectr_atm}).
Note that our model fitting of the X-ray characteristics requires a best-fit $\zeta\sim 66^\circ$ in contrast to that
found by modelling radio observations ($\zeta\sim 84^\circ$). This discrepancy will be addressed in the
Discussion.
Interestingly, we verified that adding a hot core to the cooler component in the antipodal spot
does not make
a noticeable difference to the modelled pulse profile and pulsed fraction. This is because the beam/pulse profile of the 
extra antipodal hot spot is very narrow, and given the geometry, we are not seeing the hot core of the antipodal cap.
In other words, if the two polar caps have the same temperatures (but somewhat different sizes), then 
this agrees with the X-ray observations.

\section{Summary}

Our long \xmm\ X-ray and GMRT/LOFAR radio observations of \psr\ revealed, after \psrb, a second example of simultaneous
X-ray/radio moding, and we could study its X-ray characteristics in great detail. 
First, we will present a summary of the new findings of our X-ray radio campaign.

(i) We found \psr\ during five out of our six \xmm\ observations to be in the radio B mode and during only one observation, on April 24, 2017, in the Q mode. Over our entire
radio coverage with GMRT and LOFAR of $\sim$80 hours, \psr\ spent only $\sim$15\% of the time in Q mode.

(ii) We discovered correlated X-ray/radio moding: during the five X-ray observations with \psr\ in the radio B mode, the pulsar is detected in the skymaps with an average count rate of
($10.81 \pm 0.44) \times 10^{-3}$ cnts/s (0.2$-$2 keV), and during the single radio Q mode it is not detected down to a 2$\sigma$ upper limit of $1.22 \times 10^{-3}$ cnts/s. 

(iii) A new manifestation of X-ray variability of pulsars was discovered in terms of a variation in the average X-ray flux in the five B-mode observations. The average flux varied by $\pm$20\%
around the average value. 

(iv) In the April-24 observation \psr\ started in the radio Null mode, then spent 6$-$7 minutes in the Q-bright mode, and then gradually faded to the
Q mode in which it stayed for 7 h and 50 minutes. 

(v) The X-ray behaviour was correlated with the successive radio modes during the April-24 observation. During the initial radio Null mode \psr\ was not detected in X-rays, but 
during an interval of 21 minutes including the Q-bright mode a strong (4.5$\sigma$) detection was made in the spatial analysis which then appeared to decay over $\sim$1.5 hours
to non-detectable X-ray emission in the remaining 7 hours of the Q mode. 

(vi) In the radio B mode, we discovered X-ray pulsations from \psr\  between 0.2 and 2 keV with a broad sinusoidal pulse. The X-ray pulse 
significantly lags the radio MP by $0.208 \pm 0.012$ in phase. There is no pulse visible around the phase of the IP, but a weak sinusoidal 
X-ray IP can be hidden under the minimum level of the light curve. In the Q mode we found no indication of a pulsed signal in our X-ray data. 

(vii) The X-ray pulsed fraction we found to be very high, $\sim$70-80\% for energies between 0.2 and 2 keV.

(viii) The spectrum of the total X-ray emission is well described by a double blackbody model (goodness of fit probability $\sim$30\%) 
as well as by a blackbody plus power-law model ($\sim$10\% probability). The double-BB model
has a cool component with {\it kT} $\sim$0.12 keV ($\sim$1.4 $\times 10^6$ K) with hot spot radius 
of $\sim$77 m, and a hotter blackbody with $kT \sim 0.37$ keV ($\sim$4.3 $\times 10^6$ K) and radius $\sim$8.5 m. For the BB-plus-PL model the parameters were: 
$kT \sim$0.29 keV ($\sim$3.4 $\times 10^6$ K) with hot spot radius 
of $\sim$11 m, and PL index $\Gamma \sim -2.50$.

(ix) The spectrum of the pulsed emission is also well fit with a double-BB model (goodness of fit probability $\sim$10\%) with 
the same parameters as for the total spectrum, and also with a BB-plus-PL
model ($\sim$5\% probability) with a somewhat lower temperature $kT \sim 0.18$ keV, but the uncertainties on hot spot radius and PL index become large.

(x) The spectrum of a complementary unpulsed emission, or of an Interpulse, is not well constraint due to the low counting statistics available.

(xi) Alternatively, the total spectrum of \psr, its pulse profile and pulsed fraction as a function of energy can all be reproduced with 
a magnetized partially ionized hydrogen atmosphere model
with three emission components, a primary spot with a small hot core ($kT_2 \sim 310\mbox{ eV}$ or $\sim$3.6 $\times 10^6$ K, $R_2 \sim 17\mbox{ m}$)
and a larger cooler concentric ring ($kT_1 \sim 97\mbox{ eV}$ or $\sim$1.1 $\times 10^6$ K, $R_2 \sim 280\mbox{ m}$)
and an antipodal hot spot ($kT_3 \sim 97\mbox{ eV}$, $R_3 \sim 100\mbox{ m}$). However, this atmosphere-model fit to 
our X-ray data requires the angle  $\zeta$ between the rotation axis and line of sight direction 
to be $\sim 66^\circ$, in conflict with the accurately radio-derived value of $\sim 84^\circ$.

(xii) Adding in the atmosphere-model fit a fourth emission component, namely a hot core in the antipodal cap such 
that the two polar caps have the same temperatures (but somewhat different sizes), would also
be in agreement with the X-ray observations.

 \section{Discussion}
 
 \subsection{X-ray moding: PSRs B0943+10, B1822$-$09 and B0823+26}

To date, long simultaneous X-ray/radio campaigns on radio-mode-switching pulsars have only been performed on 
\psrb\ \citep{hermsen2013, mereghetti2016}, \psrc\ \citep{hermsen2017} and on \psr, presented in this work. The discovery of synchronous X-ray/radio mode switching
was reported for the nearly aligned  \psrb: its X-ray flux switched in {\it anti correlation} with the radio flux: when the 
pulsar is in a radio ``bright'' (B) mode the X-ray flux is weak, and when it switches to a radio ``quiet'' (Q) mode, the X-ray flux more than doubles \citep{hermsen2013}.
In both radio modes the X-ray emission is the sum of a pulsed thermal component with $T_{h} \sim3.4 \times 10^6$ K (hot spot radius $R_{h}\sim$ 21 m) plus
 an about equally strong unpulsed non-thermal component with  photon index $\sim -2.5$ \citep{mereghetti2016}. 
However, for the nearly orthogonal \psrc\ no indication was found for synchronous X-ray/radio moding and its
total X-ray spectrum could be fitted with two black bodies, a cool component ($ T_{c}\sim1.0 \times 10^6$ K, $R_{c}\sim$ 2.0 km) 
and a hot component ($T_{h}\sim2.2 \times 10^6$ K, $R_{h}\sim$100 m) \citep{hermsen2017}. 
In this work on the nearly orthogonal \psr, we report the discovery  of a second example of synchronous X-ray/radio mode switching.

\begin{table*}
\caption{ Comparison of radio and X-ray characteristics of the only mode-switching pulsars that are so far observed simultaneously in the radio and X-ray bands. The errors on the
parameters derived in the X-ray analyses are for \psr\ given in the previous sections, and for \psrb\ and \psrc\ by \citet{mereghetti2016} and \citet{hermsen2017}, respectively.}
\label{Comparison}
\begin{center}

\begin{tabular}{cccc}
\hline
  &  \psr &   \psrb  &   \psrc  \\
\hline
$P$ [s]&  0.53   &   1.1    &  0.77 \\
$\dot{P}$ [s s$^{-1}$]   &   $1.7 \times 10^{-15}$  & $3.5 \times 10^{-15}$   & $5.3 \times 10^{-14}$   \\
  $\tau$[Myr] &  5 & 5  & 0.2 \\
$B$ [G]   &  $ 9.6 \times 10^{11} $ & $4 \times 10^{12}$   & $ 6.5 \times 10^{12}$  \\
$\dot{E}_{rot}$ [erg s$^{-1}$]   &      $4.5 \times 10^{32}$    &  $ 10^{32}$    &  $4.6 \times 10^{33}$ \\
Distance [pc]& 320$^i$ & 630$^{ii}$  &  900$^{iii}$ \\
Geometry &     nearly orthogonal   & nearly aligned   &  nearly orthogonal \\
Radio-mode duration &  $ \sim$hours   &  $\sim$ hours     &   $\sim$minutes \\
  &   &   &   \\
Radio X-ray moding  &   correlated  &  anti correlated     & none  \\
X-ray pulse profile & sinusoidal & sinusoidal  &  sinusoidal  \\
X-ray pulsed fraction*  & B: $\sim$80\% & B: $\sim$35\%  &  $\sim$15\% $-$ $ \sim$60\% \\
 &  Q: no detection   &  Q: $\sim$20\% $-$ $\sim$65\%  & \\    
Total X-ray spectrum &  BB$_1$+BB$_2$** &   BB$_1$+PL (index -2.5)  &   BB$_1$+BB$_2$ \\
  $ T$ $_{BB_1}$[K] &  1.4 $\times 10^6$  & 3.0 $\times 10^6$&     1.0 $\times 10^6$\\
   $R_{BB_1}$ [m] &   77 &    21 &  2000 \\   
   $T$ $_{BB_2}$[K]    &  4.3 $\times 10^6$ &    &  2.2 $\times 10^6$ \\
   $R_{BB_2}$ [m] &   8.5 &     &   100 \\ 
$L_X$ Total [erg s$^{-1}$]  & $1.6 \times 10^{29}$   &      &  $3.7 \times 10^{30}$ \\
$L_X$ B mode [erg s$^{-1}$]  &    & BB: $ 1.9 \times 10^{29}$  &    \\
   &   &   PL: $0.8 \times 10^{29}$ & \\
   $L_X$ Q mode [erg s$^{-1}$]  &    & BB: $2.3 \times 10^{29}$ &    \\
      &   &   PL:  $3.5 \times 10^{29}$ & \\
  Pulsed X-ray spectrum***  & BB+BB** &   BB  &   BB+BB \\
$L _X$ Pulsed Total [erg s$^{-1}$] &   &  &   $1.0 \times 10^{30}$ \\  
$L _X$ Pulsed B mode [erg s$^{-1}$] &  $1.1 \times 10^{29}$ & $1.0 \times 10^{29}$ &    \\
 $L _X$ Pulsed Q mode [erg s$^{-1}$]  &    &       $ 2.9 \times 10^{29}$ &  \\

   \hline
 \multicolumn{4}{l} {$^i$ \citet{verbiest2012}; $^{ii}$ based on a pulsar dispersion measure of 15.32 pc cm$^{-3}$ \citep{bilous2016}} \\
 \multicolumn{4}{l} {and the \citet{cordes2002} galactic electron density distribution;  } \\ 
 \multicolumn{4}{l}{a more recent electron density model \citep{yao2017} gives a larger distance of 890 pc; }\\ 
 \multicolumn{4}{l}{$^{iii}$ \citet{zou2005};  an upper limit to the distance of 1.9 kpc is given by \citet{verbiest2012}.} \\
\multicolumn{4}{l} {*Average value over energy interval 0.2 $-$ 2 keV or values at start and end of this energy interval; }\\
\multicolumn{4}{l}{**Less likely fit with BB ($T$ = 3.0 $\times 10^6$ K) + PL (index -2.5). }\\
\multicolumn{4}{l}{***For definition pulsed emission see Table 4: {\it Pulse-ME}. All luminosities $L_X$ for energies 0.5$-$2 keV. }\\
\end{tabular}
\end{center}
\end{table*}


Most surprisingly, the characteristics of the X-ray moding and X-ray emission of \psr\ are very different from those reported for \psrb.
In particular, we revealed {\it correlated} X-ray/radio moding: in the radio B mode \psr\ is detected in X-rays and exhibits a dominant pulsed component,
while the pulsar is not detected during the radio Q mode, with a 2{$\sigma$} flux upper limit a factor $\sim$9 below the X-ray flux in the B mode.
We showed that the total emission of \psr\ can also be described by a black-body and power-law component with the same temperature and index
as measured for \psrb. The difference being that for \psr\ both components contribute to the pulsed emission, while the pulsed emission of \psrb\ is
purely thermal. However, the statistically preferred model spectrum for \psr\
consists of two black bodies, a hot (small hot spot radius $R_{h}\sim$8.5 m) component 
with $ T_{h} \sim4.3 \times 10^6$ K and a cool (broader $R_{c}$ of $\sim$77 m) component with $T_{c} \sim1.4 \times 10^6$ K.
In our discussion we will address this latter preferred spectral shape. 

The aim of our campaign was to further our understanding of X-ray/radio mode switching. However, the emerging picture is becoming
increasingly more complex. The three pulsars have different geometries (orientations of magnetic and rotation axes and viewing angles), 
different X-ray spectral shapes and/or different mode-switching characteristics
(mode durations and correlated/anti correlated or no X-ray/radio mode switching).
On the other hand, for all three pulsars we detected broad sinusoidal X-ray pulse profiles with (very) high pulsed fractions between $\sim$60\% and $\sim$80\% around 1 keV.
Table~\ref{Comparison} summarises the radio and X-ray characteristics. In the discussion we will address the different moding characteristics,
the differences in X-ray spectral and timing properties, as well as the enigmatic problem of what might cause mode switching in pulsars.

\subsection{X-ray mode-switch time scales}

\psrb\ and \psr\ are both old pulsars ($\tau \sim$ 5 Myr) and exhibit mode durations of the order of hours, while the much younger \psrc\ ($\tau \sim$ 0.2 Myr) 
switches radio modes on time scales of minutes. Within the uncertainties of radio single-pulse studies, 
radio-mode switches are seen to occur on time scales of single rotations.
In our X-ray data we can not accurately study mode-switch time scales, because these pulsars are 
weak X-ray sources from which we measure with \xmm\ count rates of order 0.01 cnts s$^{-1}$.
An attempt was made for \psrb: the X-ray mode transitions were studied in X-ray light curves 
obtained by {\it stacking} all the Q to B and B to Q mode transitions observed in a campaign in 2014 (Fig. 6 in \citet{mereghetti2016}).
Within the available statistics it was concluded that the X-ray transition occurred within 15 minutes after the radio switch. 
In Fig.~\ref{April24ZOOM}, we show for a {\it single} transition of \psr\ that the X-ray emission was
significantly detected in a time interval of 21 minutes after the switch from a null mode to the short radio Q-bright mode on 24 April. 
The low \xmm\ counting statistics of this weak X-ray source do not allow us to study single X-ray mode switches on shorter time scales even though the \xmm\ time
resolution is 47.7 ms. 
Therefore, we can not exclude that the explanation for the non detection of X-ray moding of \psrc\ is that the minutes 
long mode durations of \psrc\ are too short to allow the pulsar to react in X-rays to a radio-mode switch.

In Fig.~\ref{PSRB0823lightcurve} we presented the \psr\ count rates (0.2$-$2 keV) in time intervals of $\sim$1 hour
covering all six observations of our campaign in April 2017. This figure revealed the \psr\ X-ray variability due to radio-mode switching, but also
that the X-ray count rate in the B-mode is not constant. This is a new phenomenon. X-ray emission from radio pulsars was for decades considered
to be stable, until the discovery of synchronous X-ray/radio mode switching. Also, two high-B pulsars
exhibited magnetar-like outbursts, PSR J1846$-$0258 \citep{gavriil2008} and PSR J1119$-$6127  \citep{antonopoulou2016, archibald2016}, but what we see here is different.
There appears to be some variability in the X-ray flux of order $\pm$20\% on time scales of hours. E.g. the average X-ray flux during our 9-hours observation
on 20 April was $\sim$40\% higher than during our 7-hours observation on 30 April. 

When we examined our GMRT monitoring data to identify the radio modes, 
we noticed that in the B mode, the pulsar is bright and displays short (a few pulses) nulls. Also, on 28 April 2017 during the B mode there are short durations of about
200 pulses of weak emission, but different from the Q state, and not classified as a separate mode. 
If the frequency and lengths of short nulls or the short durations of weak emission during the B mode varies, such as recently discussed for PSR J1822$-$2256 \citep{basu2018}, and
these occurred e.g. more frequently in our observation on 20 April compared to 30 April, then this might offer an explanation for the variations in average X-ray count rates.
However, if that is the explanation, it would also mean that the X-ray flux is closely correlated with the radio flux on time scales of possibly even a few pulses. 
If that would apply for \psrc\ as well, then mode switching in its X-ray emission could also have been detected for its short mode durations of minutes.  A detailed single-pulse study
of our radio observations of \psr\ is required to investigate such a possible tight X-ray/radio correlation, but it is beyond the scope of this work and will be addressed in a follow-up investigation.
 
\subsection{X-ray characteristics: broad thermal pulses, high pulsed fractions from hot polar caps}

We noted above that all three pulsars exhibit broad sinusoidal pulse profiles with very high pulsed fractions. 
\psrb\ has a pulsed fraction in the Q mode that increases with energy from $\sim$20\% at 0.2 keV to $\sim$65\% at 2 keV, 
while the X-ray pulsed fraction in the B mode ($\sim$ 38\%) did not vary with energy \citep{mereghetti2016}.
For \psrc\  similar pulsed fractions were reported: $\sim$15\% around 0.3 keV up to $\sim$60\% at 1 keV \citep{hermsen2017}, 
and for \psr\ we find in the B mode even higher pulsed fractions ($\sim$70$-$80\% for energies 0.2$-$2 keV).
Furthermore, for all three pulsars the pulsed emissions are thermal. Namely, a single black-body spectrum for \psrb, and 
double-black-body spectra for \psr\ and \psrc. In addition, the derived values for $R_{BB}$ are smaller than
the radius of the polar cap, defined by the last closed lines in a dipolar field geometry, $ R_{PC} = ((2\pi R^3)/(Pc))^{1/2}$, 
with the exception of the cool component of \psrc. 
For the latter pulsar, which is more than an order of magnitude younger and has a rotational energy loss an order of 
magnitude larger (see Table~\ref{Comparison}), $R_{BB_c}$ = $\sim$2 km, compared to 
a polar-cap radius of 165 m (assuming $R$=10 km). This cooler component might be related to thermal emission from the entire surface. 

Thus the hot thermal components of \psrb, \psrc\ and \psr, as well as the cooler component of \psr, have $R_{BB}$ smaller than their polar cap radii. 
These polar hot spots are explained as the 
result of bombardment of the polar cap by electrons accelerated along the last closed field lines towards the neutron star.
However, such high pulsed fractions can not be explained with models in which thermal emission is isotropically emitted from hot polar caps. 
Including gravitational light bending, it has been shown
that the pulsed fraction can not exceed 30\% \citep{pechenick1983, beloborodov2002, bogdanov2007, bogdanov2016}.  
This is the case for nearly orthogonal geometries (\psr\ and \psrc), 
as well as for nearly aligned pulsars (\psrb). 

In Section \ref{atmosphere-model}, we showed that a more physically correct atmosphere model can reproduce 
very high pulsed fractions up to $\sim$ 90\% for pulsars with sufficiently strong magnetic fields. 
For modelling the case of the nearly orthogonal \psr, with both poles visible every rotation, we used a magnetized 
partially ionized hydrogen atmosphere model 
and took relativistic effects into account \citep{ho2007}. The X-ray spectrum, pulse profile and pulsed fractions of \psr\ 
were shown to be approximately matched (see Fig.~\ref{spectr_atm}) with a model that  includes three emission 
components: a primary spot with a small hot core ($T_{h1} = 3.6 \times 10^6 $ K, $R_{h1}$ = 17 m) 
and a larger cooler annular ring ( $T_{c1} = 1.1 \times 10^6$ K, $R_{c1}$ = 280 m), and an antipodal `cool' 
spot ($T_{c2} = 1.1 \times 10^6$ K,  $R_{c2}$ = 100 m). Adding an extra antipodal hot spot in the model does not make a significant difference
in the predicted X-ray characteristics, thus, interestingly, if the two polar caps have the same temperatures
(but somewhat different measured sizes), this also agrees with the X-ray data. It is interesting to note that 
the radius for the cool primary component (280 m) is similar to
 the polar cap radius: $R_{PC}$ is 199 m for R = 10 km and 295 m for R = 13 km.

\psrb\ is a rotator that is viewed under an angle of only 9$\degr$ \citep{deshpande2001}. For such a geometry one 
polar cap is viewed continuously, and one would expect to observe a practically flat
pulse-phase distribution if the thermal emission originates in an isotropically emitting hot spot. \citet{storch2014} 
showed that also for this geometry of \psrb\ pulsed fractions close to 60\% (see also \citep{mereghetti2016})
can be explained by including the beaming effects of a magnetised atmosphere, while remaining consistent with 
the dipole field geometry constrained by radio observations. The latter is, unfortunately, not the case
for \psr. The successful modelling of the X-ray characteristics of \psr\ as shown in Fig.~\ref{spectr_atm}, 
was realised for values of $\alpha$ = 81$^{\circ}$, the angle between the magnetic and rotation axes, and by
$\zeta$ = 66$^{\circ}$, the angle between the rotation axis and the line of sight.  An acceptable solution can also 
be obtained for $(\alpha,\zeta) = (66^\circ,81^\circ)$. We noted that fits to radio polarisation data using the rotating vector
model give $(\alpha,\zeta) = (81^\circ,84^\circ)$, with an uncertainty of $0.7^\circ$ \citep{everett2001}. 
Therefore, in either case there is an inconsistency between the radio and X-ray derived geometries. 
Furthermore, the broad X-ray pulse is lagging the radio main pulse in phase by as much as 0.208 $\pm$ 0.012. 
In order to explain such a large phase shift, one might have to invoke a displaced dipole geometry or a non-dipolar magnetic field close
to the surface where the X-rays are produced. Nevertheless, it is not obvious how to relate the phase shift
to the different geometrical angles derived from the X-ray and radio characteristics.

In Section~\ref{Spectrum total} we concluded that with lower probability (goodness of fit 10\%) the X-ray spectrum of \psr\ can also be 
described as the sum of a black body ($T$ $\sim 3.4 \times 10^{6}$ K, $R_{PC} \sim11$ m) and a power law (index $-2.5$),
with the same spectral parameters as derived for \psrb. For the latter pulsar, its pulsed emission is thermal (for a 
possible explanation, see above), and its unpulsed component is non-thermal. For its nearly aligned geometry the  
unpulsed emission could be non-thermal beamed emission, continuously viewed under the small viewing angle. 
Could such an explanation also work for the nearly orthogonal \psr? In that case the pulsed emission is the sum of a thermal component
plus a non-thermal component, {\it both} contributing to the broad pulse profile.  
Thus the thermal and non-thermal emissions should both be beamed in the same direction and both beams should be broad. 
Alternatively, the beams are narrow and shifted in phase by e.g. 0.2, together forming the measured broad profile. In the latter
case the shape of the pulse profile should be energy dependent. However, in Fig.~\ref{ProfilesB0823} the profile shapes in 
the energy intervals 0.2$-$0.5 and 0.5$-$2 keV are statistically identical. Furthermore,
for a nearly orthogonal geometry and a viewing angle close to 90$^\circ$, one would then expect to see beams of non-thermal emission from both poles.

\subsection{The nature of mode changing in \psr}

The results of our campaign are difficult to interpret consistently in terms of current ideas about the nature of polar cap physics 
and to reconcile the directions of the radio and X-ray emission beaming in \psr\ (Sections 6.3 and 8.3). It is therefore useful 
to stand back and ask more fundamental questions about the nature of mode-changing in this pulsar. Our clear result that the 
pulsed X-rays of \psr\ switch on and off in tandem with the radio mode switch stands in stark contrast to the result of our 
earlier campaign on \psrc\ in which no detectable X-ray mode switch was found. Both pulsars are near-perpendicular 
rotators and exhibit clearly-defined modes designated as B(right) and Q(uiet). So why are they so different?  

The most obvious and immediate difference between these pulsars is the timescale and relative intensity of their radio-defined modes. 
In \psrc\ the separate modes are rarely sustained for an hour and are mostly of far shorter duration, so mode changes 
occur typically every few minutes \citep{hermsen2017}. Our sustained observations here have shown that in \psr\ 
the modes appear to be much more stable, with both B and Q lasting many hours, possibly even days. 
Further, the average intensities of the B/Q modes in \psrc\ is just 2:1 \citep{fowler1981}, while in \psr\ \citet{sobey2015} 
report that the Q mode is on average a factor $\sim$200 dimmer that the B mode, such that earlier workers had assumed the Q mode to be a sustained null \citep{young2012}. 

The character of the modal emission in each pulsar is also very different.  In \psrc\ the onset of both the B and Q modes 
occur within a few pulses, and while there is evidence of occasional 
complex modal overlap \citep{latham2012}, there is no suggestion of one mode gradually fading into another. 
The duration of each mode has a bimodal distribution \citep{hermsen2017}, the shorter modes being of the order of 
minutes and the other tens of minutes. Each mode has been found to have a characteristic modulation timescale, 
which, in shorter mode sequences at least, seems to determine the duration of the mode. A striking feature of the 
radio observations is that the precursor suddenly becomes present in the B mode and vanishes in the Q mode and 
coincides with the disappearance of the IP, suggesting a rapid reorganisation of the entire magnetosphere \citep{goodwin2004, contopoulos2005, timokhin2010}.

The radio emission of \psr\ does not show this high degree of organisation. The onset of the B mode after the Q mode is indeed sudden, 
as in \psrc\, but the transition back to quiescence is better characterised as `flickering': a gradual lessening in the overall brightness 
of individual pulses and an increase in the number of apparently weak or null pulses. This behaviour is present on a wide range 
of timescales. It is found  in our observations on April 24 after the 7-minute appearance of the Q bright mode 
(Fig.~\ref{April24GMRT306MHz}), in the even briefer 1.4 minute long B-mode-like emission reported by \citet{sobey2015} and 
also at the conclusion of hours-long B mode sequences  \citep{rankin2018}. Furthermore, in contrast to \psrc, no switching 
in the presence/absence of the PC or IP is found in the Q mode other than a reduction in intensity of these components 
roughly in proportion to the reduction in the MP intensity \citep{rankin2018}. 

Unfortunately no B-to-Q transitions were observed in our simultaneous X-ray and radio campaign, but it may be significant 
that the X-rays coinciding with the Q-bright mode of April 24 take about 100 minutes to diminish although only sporadic 
radio emission is detected (Fig.~\ref{April24ZOOM}). It is possible that this happens at the end of every B mode sequence, 
indicating that radio emission is still occurring but undetectable. This would support the reasonable conjecture (see Section 8.2) 
that X-ray power is strongly correlated to the radio emission on every timescale down to the single pulses, suggesting that 
the detection of X-rays may be used to infer the presence of unseen radio emission. The low X-ray count rate, however, 
does not allow us to follow the null/burst/flicker pattern. Since in the radio emission this pattern is found on a wide range 
of timescales it is tempting to assume that the X-ray emission follows suit. Perhaps the reduction in the B mode 
X-rays during our on average 7-hours long observations (Fig.~\ref{PSRB0823lightcurve}), and likely also visible in the observation 
by \citet{becker2004} (Fig.~\ref{Beckerlightcurve}), is also an indicator of the radio pattern existing on an even longer timescale.

A further difference between the radio behaviours of the two pulsars can be seen in the timescales and nature of 
their single-pulse modulation $P_3$. In \psrc\ the modulation timescales in both modes (Q mode $P_3$$\sim$46$P$ and B mode $\sim$70$P$ 
 \citep{latham2012}) can be seen as roughly compatible with the predicted value of the \citet{ruderman1975} 
carousel model and the emission is null-free. By contrast, \psr\ exhibits a B-mode modulation of $P_3$$\sim$5$P$ 
that is significantly shorter, and, as noted above, appears as a burst rather than a simple smooth cycle. 
This was originally pointed out by \citet{romani1992}, who calculated a low fractal dimension for the attractor present 
in the chaotic null/burst behaviour. The modulation may therefore have a different physical origin to the carousel-based  
`drift' phenomenon found in other (often near-aligned) pulsars such as \psrb. It is not impossible that the burst pattern 
persists in the Q mode but the weakness of the radio emission makes it undetectable.  

It is interesting to speculate that in \psr\ we are not seeing `true' mode-changing but the sudden appearance of 
strong bursts whose intensity follows a self-similar (\ie fractal) distribution over a wide range of timescales. 
Thus what we identify as Q mode is simply emission largely hidden by our inability to detect it. Such systems are sometimes 
identified as exhibiting self-organised criticality (SOC) found in non-linear slowly-driven non-equilibrium systems such as 
avalanche theory \citep{bak1987} and may underlie the Rotating-Radio-Transient (RRAT) phenomenon \citep{mclaughlin2006} that the Q mode strongly resembles 
\footnote{Not all systems reach a high degree of criticality: in an investigation of the time series of 17 pulsars only 
one (PSR B1828$-$11) was found to possess clear chaotic behaviour \citep{seymour2013}.}.

If this picture is correct, then the sudden switches between two well-defined and null-free modes found in \psrc\ 
and other pulsars (including \psrb) are a very different phenomenon to what we find in \psr. However, 
caution is necessary since SOC systems are often characterised by power-law distributions, and while this was 
found to be the case in the weak Q-mode distributions analysed by \citet{sobey2015}, these authors found 
the B mode emission to be better modelled by lognormal statistics. Clearly further study of our longer radio observations could be used to investigate this.  

Some years ago it was pointed out by \citet{weltevrede2006b} that the middle-aged (X-ray detected) pulsar 
PSR B0656+17 simultaneously exhibited two types of radio emission, one of which was `spiky' and would 
have been observed as RRAT emission if the pulsar was located at a greater distance. This implies the risk that the level of 
radio detectability determines the classification of the pulsar's apparent mode rather than its intrinsic nature. 
Pursuing this argument, it is interesting to compare \psr\ to PSR J1752+2359, a pulsar which bears a close resemblance to \psr\ both in its emission 
behaviour and basic parameters (P=0.4s,  $\tau\sim$10 Myr, weak magnetic field $\sim$5$\times10^{11} $ G), 
but is only in its burst state for little more than 10\% of the time. The burst sequences have a structure similar to those 
described above for \psr\ and occasional strong inter-burst pulses are found to appear at random among the 
long `nulls' (see Figures 7 and 11 of \citet{gajjar2014}). However, this pulsar is located ten times further away than 
\psr\ (3 kpc compared with 0.3 kpc) which may simply make many bursts impossible to detect.

If \psr\ is a different kind of pulsar to \psrc, how do we account for this physically? A pulsar is generally 
considered to exist in a vacuum and its diverse radio phenomena, including mode changing, result entirely from 
non-linear interactions between the magnetosphere and the polar cap. This may well be the case for \psrc.  
However, the self-similar behaviour of \psr\ over many timescales, if it can be confirmed, suggests a different scenario. 

The best-known example of a SOC system displays avalanches of different magnitudes 
being triggered by a light but steady rain of sand on a sandpile \citep{bak1987}. By analogy, it is interesting to speculate 
that \psr\ is accreting material either from the interstellar medium through which it is passing, or, more specifically
from fallback debris created by its own supernova.
 The idea of accretion 
by apparently isolated pulsars is not new \citep{wright1979, tsygan1980, cheng1985, luo2007, cordes2008} 
and supposes that neutral hydrogen can pass through the bow shock of the extended magnetosphere and 
infiltrate the light cylinder where it becomes ionised and interferes with the pulsar's emission either as an
intermittent stream or via a disk. 
The former is a similar idea to the impact of the solar wind on the terrestrial ionosphere, leading to auroral effects, 
which indeed have been shown to follow SOC statistics (see review \citet{aschwanden2016}). 

If sufficient material is available, e.g. from supernova fallback material, it can give rise to a neutral accretion disk \citep{cordes2008} and 
in the context of \psr\ may help to explain the asymmetry of the X-ray pulse 
as being a combination of hot polar cap emission and cooler but more extended equatorial heating arising 
from the down flow of ionised material generated by the passing polar beam. The presence of an accretion 
disk may also account for the partial obscuration of the emission. Clearly, these ideas need to be worked out more carefully 
and they are presented to resolve the impasse between our observational results and existing polar cap theory.

\section{Conclusions}

We observed the radio-mode switching \psr\ for $\sim$39 hours simultaneously in X-rays and the radio band and report  
the discovery of synchronous {\it correlated} X-ray \& radio mode switching:
when \psr\ is radio bright (B), we detect the pulsar in X-rays in the energy band 0.2$-$2 keV with a high pulsed fraction of 70$-$80\% 
due to a sinusoidal pulse that is lagging the radio main pulse by $\sim$ 0.2 in phase. 
During the radio-quiet or null mode (Q) we do not detect \psr\ in X-rays with an upper limit almost an order of magnitude
lower than the reported flux in the B mode. This result is surprising, because the only pulsar for which synchronous X-ray \& radio moding
was found so far, \psrb, showed {\it anti-correlated} mode switches \citep{hermsen2013}. The latter pulsar exhibits thermal pulsed as well as non-thermal unpulsed X-ray emission,
both varying by a factor $\sim$ 2 in flux between similar radio B and Q modes \citep{mereghetti2016}. 

\psrb\  is classified as a nearly aligned rotator (near alignment of the magnetic and rotation axis) and \psr\ as a near-perpendicular rotator,
but these pulsars have in common their broad thermal X-ray pulses with high pulsed fractions.  Such high pulsed fractions can not be
generated modelling black-body emission isotropically emitted from a hot polar cap. For \psr, we showed that its X-ray spectrum, pulse shape and 
pulsed fraction can de reproduced with a magnetized partially ionized hydrogen atmosphere model, with the two polar caps having the same temperatures
but somewhat different sizes.  However, this atmosphere-model fit to 
our X-ray data requires the angle between the rotation axis and line of sight direction 
to be $\sim66^\circ$, in conflict with the accurately radio-derived value of $\sim84^\circ$. \citet{storch2014} applied a similar atmosphere model to the X-ray data of \psrb\ 
and approximately reproduced also for this nearly aligned pulsar the pulse profile and high pulsed fraction, but in this case the radio-derived angles could be used in the model.
In conclusion, we do not have an explanation yet for the high X-ray pulsed fractions that is consistently in agreement with radio-derived parameters.

In this work we discovered, in addition to the synchronous X-ray \& radio mode switching, a new type of X-ray variability: the average X-ray flux within B-mode intervals of duration $\sim$ 7 hours
varied by a factor $\pm$20\%. We speculate that the X-ray generation follows very closely the variations in frequency and lengths of short radio nulls or short durations of weak emission seen during the B mode,
possibly on time scales of even few pulses. This possibility offers interesting constraints on the interpretation of what is causing mode switching: are we dealing with a local or global magnetospheric phenomenon?
A follow-up investigation of the available single-pulse radio data is required to investigate such a possible tight X-ray/radio correlation.

In our discussion on the possible nature of mode changing in \psr, we discussed in detail the radio and X-ray characteristics of \psr\ compared to those of the also radio-mode switching \psrc. For the latter pulsar an earlier X-ray/radio campaign did not
reveal synchronous moding \citep{hermsen2017}. Both pulsars are near-perpendicular rotators. We are speculating that in \psr\ we are not seeing `true' mode-changing but the sudden appearance of 
strong bursts whose intensities follow a self-similar (\ie fractal) distribution over a wide range of timescales. Such a system could be identified as exhibiting self-organized criticality. In this context, we speculate that 
\psr\ is accreting material from a debris disk or the interstellar medium through which it is passing, to explain some of its X-ray characteristics. Further study of rotation-powered radio
pulsars using simultaneous X-ray/radio data is needed to test and develop these various hypotheses.

\section*{Acknowledgments}

We thank the staff of {\it XMM-Newton}, GMRT, LOFAR and Lovell for making these observations possible. {\it XMM-Newton} is an ESA science mission
with instruments and contributions directly funded by ESA member states and by NASA. GMRT is run by the National Centre for Radio Astrophysics of the Tata
Institute of Fundamental Research. This paper is based (in part) on results obtained with International 
LOFAR Telescope (ILT) equipment, as part of project LC7-014. LOFAR is the Low Frequency Array designed and constructed 
by ASTRON. It has observing, data processing, and data storage 
facilities in several countries, that are owned by various parties (each 
with their own funding sources), and that are collectively operated by 
the ILT foundation under a joint scientific policy. The ILT resources 
have benefitted from the following recent major funding sources: 
CNRS-INSU, Observatoire de Paris and Universit\'{e} d'Orl\'{e}ans, France; BMBF, 
MIWF-NRW, MPG, Germany; Science Foundation Ireland (SFI), Department of 
Business, Enterprise and Innovation (DBEI), Ireland; the Netherlands Organisation for Scientific Research (NWO), The 
Netherlands; The Science and Technology Facilities Council, UK7.
Pulsar research at the Jodrell Bank Centre for Astrophysics and the observations using the Lovell telescope is 
supported by a consolidated grant from the STFC in the UK. 
We acknowledge the use of the international LOFAR stations operated by the MPIfR (DE601), the Nan\c{c}ay Radio Observatory, operated
by Paris Observatory, associated with the French Centre
National de la Recherche Scientifique and Universit\'{e}
dÕOrl\'{e}ans (FR606).
ASTRON and SRON are supported financially by NWO.
JMR acknowledges funding from a NASA Space Grant.
JWTH acknowledges funding from an NWO Vidi fellowship and from the European Research Council under the European Union's 
Seventh Framework Programme (FP/2007-2013) / ERC Starting Grant agreement nr. 337062 (``DRAGNET").
WCGH acknowledges support from STFC in the UK through grant number ST/M000931/1, and appreciates use of computer facilities
at the Kavli Institute for Particle Astrophysics and Cosmology.
GW thanks the University of Manchester to granting Visitor status.
DM acknowledges funding from the grant ``Indo-French Centre for the Promotion of Advanced Research, CEFIPRA''.

\bibliographystyle{mnras}
\bibliography{pulsars}

\label{lastpage}

\end{document}